\documentclass[longauth]{aa}
\usepackage[utf8]{inputenc}
\usepackage{array}
\begin{document}

\title{The \emph{Gaia}-ESO Survey: properties of newly discovered Li-rich giants\thanks{Based on observations collected at the European Organisation for Astronomical Research in the Southern Hemisphere under ESO programmes 188.B-3002 and 193.B-0936 (The \emph{Gaia}-ESO Public Spectroscopic Survey)}}

\author{
R. Smiljanic\inst{1}
\and
E. Franciosini\inst{2}
\and
A. Bragaglia\inst{3}
\and
G. Tautvai{\v s}ien{\.e}\inst{4}
\and
X. Fu\inst{3,5}
\and
E. Pancino\inst{2,6}
\and
V. Adibekyan\inst{7}
\and
S.~G. Sousa\inst{7}
\and
S. Randich\inst{2}
\and
J. Montalb\'an\inst{8}
\and
L. Pasquini\inst{9}
\and
L. Magrini\inst{2}
\and
A. Drazdauskas\inst{4}
\and
R.~A. Garc\'{\i}a\inst{10,11}
\and
S. Mathur\inst{12,13,14}
\and
B. Mosser\inst{15}
\and
C. R\'egulo\inst{12,13}
\and
R. de Assis Peralta\inst{15}
\and
S. Hekker\inst{16,17}
\and
D. Feuillet\inst{18}
\and
M. Valentini\inst{19}
\and
T. Morel\inst{20}
\and
S. Martell\inst{21}
\and
G. Gilmore\inst{22}
\and
S. Feltzing\inst{23}
\and
A. Vallenari\inst{24}
\and
T. Bensby\inst{23}
\and
A.~J. Korn\inst{25}
\and
A.~C. Lanzafame\inst{26}
\and
A. Recio-Blanco\inst{27}
\and
A. Bayo\inst{28,36}
\and
G. Carraro\inst{8}
\and
M.~T. Costado\inst{29}
\and
A. Frasca\inst{30}
\and
P. Jofr\'e\inst{31}
\and
C. Lardo\inst{32}
\and
P. de Laverny\inst{27}
\and
K. Lind\inst{18,25}
\and
T. Masseron\inst{12,13}
\and
L. Monaco\inst{33}
\and
L. Morbidelli\inst{2}
\and
L. Prisinzano\inst{34}
\and
L. Sbordone\inst{35}
\and
S. Zaggia\inst{24}
}

\institute{Nicolaus Copernicus Astronomical Center, Polish Academy of Sciences, Bartycka 18, 00-716, Warsaw, Poland\\ \email{rsmiljanic@camk.edu.pl}
\and
INAF - Osservatorio Astrofisico di Arcetri, Largo E. Fermi 5, 50125, Florence, Italy
\and
INAF - Osservatorio di Astrofisica e Scienza dello Spazio di Bologna, via Gobetti 93/3, 40129, Bologna, Italy
\and
Institute of Theoretical Physics and Astronomy, Vilnius University, Saul{\. e}tekio av. 3, 10257 Vilnius, Lithuania
\and
Dipartimento di Fisica \& Astronomia, Universit\`a degli Studi di Bologna, via Gobetti 93/2, I-40129 Bologna, Italy
\and
Space Science Data Center - Agenzia Spaziale Italiana, via del Politecnico, s.n.c., I-00133, Roma, Italy
\and
Instituto de Astrof\'isica e Ci\^encias do Espa\c{c}o, Universidade do Porto, CAUP, Rua das Estrelas, 4150-762 Porto, Portugal
\and
Dipartimento di Fisica e Astronomia Galileo Galilei, Universit\`a di Padova, Vicolo dell'Osservatorio 3, I-35122 Padova, Italy
\and
European Southern Observatory, Karl-Schwarzschild-Strasse 2, 85748, Garching bei M\"unchen, Germany
\and
IRFU, CEA, Universit\'e Paris-Saclay, F-91191 Gif-sur-Yvette, France
\and
Universit\'e Paris Diderot, AIM, Sorbonne Paris Cit\'e, CEA, CNRS, F-91191 Gif-sur-Yvette, France
\and
Instituto de Astrof\'{\i}sica de Canarias, E-38205 La Laguna, Tenerife, Spain
\and
Universidad de La Laguna, Dept. Astrof\'{\i}sica, E-38206 La Laguna, Tenerife, Spain
\and
Space Science Institute, 4750 Walnut street, Suite 205, Boulder, CO 80301, USA
\and
LESIA, Observatoire de Paris, Universit\'e PSL, CNRS, Sorbonne Universit\'e, Universit\'e Paris Diderot, Sorbonne Paris Cit\'e,  92195 Meudon, France
\and
Max-Planck-Institut for Solar System Research, Justus-von-Liebig-Weg 3, 37077, G\"ottingen, Germany
\and
Stellar Astrophysics Centre, Department of Physics and Astronomy, Aarhus University, Ny Munkegade 120, DK-8000 Aarhus C, Denmark
\and
Max-Planck Institut f\"{u}r Astronomie, K\"{o}nigstuhl 17, 69117 Heidelberg, Germany
\and
Leibniz-Institut f\"{u}r Astrophysik Potsdam (AIP), An der Sternwarte 16, 14482, Potsdam, Germany
\and
Space sciences, Technologies and Astrophysics Research (STAR) Institute, Universit\'e de Li\`ege, Quartier Agora, All\'ee du 6 Ao\^ut 19c, B\^at. B5C, B4000-Li\`ege, Belgium 
\and
School of Physics, University of New South Wales, Sydney, NSW 2052, Australia
\and
Institute of Astronomy, University of Cambridge, Madingley Road, Cambridge CB3 0HA, United Kingdom
\and
Lund Observatory, Department of Astronomy and Theoretical Physics, Box 43, SE-221 00 Lund, Sweden
\and
INAF - Padova Observatory, Vicolo dell'Osservatorio 5, 35122 Padova, Italy
\and
Department of Physics and Astronomy, Uppsala University, Box 516, SE-751 20 Uppsala, Sweden
\and
Dipartimento di Fisica e Astronomia, Sezione Astrofisica, Universit\'{a} di Catania, via S. Sofia 78, 95123, Catania, Italy
\and
Laboratoire Lagrange (UMR7293), Universit\'e de Nice Sophia Antipolis, CNRS,Observatoire de la C\^ote d'Azur, CS 34229,F-06304 Nice cedex 4, France
\and
Instituto de F\'{\i}sica y Astronom\'{\i}a, Universidad de Valpara\'{\i}so, Chile
\and
Departamento de Did\'actica, Universidad de C\'adiz, 11519 Puerto Real, C\'adiz, Spain
\and
INAF - Osservatorio Astrofisico di Catania, via S. Sofia 78, 95123, Catania, Italy
\and
N\'ucleo de Astronom\'{i}a, Facultad de Ingenier\'{i}a, Universidad Diego Portales, Av. Ej\'ercito 441, Santiago, Chile
\and
Laboratoire d'astrophysique, Ecole Polytechnique F\`ed\`erale de Lausanne (EPFL), Observatoire de Sauverny, CH-1290 Versoix, Switzerland
\and
Departamento de Ciencias Fisicas, Universidad Andres Bello, Fernandez Concha 700, Las Condes, Santiago, Chile
\and
INAF - Osservatorio Astronomico di Palermo, Piazza del Parlamento 1, 90134, Palermo, Italy
\and
European Southern Observatory, Alonso de Cordova 3107 Vitacura, Santiago de Chile, Chile
\and
N\'ucleo Milenio Formaci\'on Planetaria - NPF, Universidad de Valpara\'iso, Av. Gran Breta\~na 1111, Valpara\'iso, Chile
}

\date{September 2017}

\titlerunning{Properties of newly discovered Li-rich giants}
\authorrunning{Smiljanic et al.}

\abstract
    % context (optional)
{}
    % aims
{We report 20 new lithium-rich giants discovered within the \emph{Gaia}-ESO Survey, including the first Li-rich giant with an evolutionary stage confirmed by CoRoT (Convection, Rotation and planetary Transits) data. We present a detailed overview of the properties of these 20 stars.}
    % methods
{Atmospheric parameters and abundances were derived in model atmosphere analyses using medium-resolution GIRAFFE or high-resolution UVES (Ultraviolet and Visual Echelle Spectrograph) spectra. These results are part of the fifth internal data release of the \emph{Gaia}-ESO Survey. The Li abundances were corrected for non-local thermodynamical equilibrium effects. Other stellar properties were investigated for additional peculiarities (the core of strong lines for signs of magnetic activity, infrared magnitudes, rotational velocities, chemical abundances, and Galactic velocities). We used \emph{Gaia} DR2 parallaxes to estimate distances and luminosities.}   
    % results
{The giants have A(Li) $>$ 2.2~dex. The majority of them (14 of 20 stars) are in the CoRoT fields. Four giants are located in the field of three open clusters, but are not members. Two giants were observed in fields towards the Galactic bulge, but likely lie in the inner disc. One of the bulge field giants is super Li-rich with A(Li) = 4.0~dex.}
%
    % conclusion (optional)
{We identified one giant with infrared excess at 22$\mu$m. Two other giants, with large $v~\sin~i$, might be Li-rich because of planet engulfment. Another giant is found to be barium enhanced and thus could have accreted material from a former asymptotic
giant branch companion. Otherwise, in addition to the Li enrichment, the evolutionary stages are the only other connection between these new Li-rich giants. The CoRoT data confirm that one Li-rich giant is at the core-He burning stage. The other giants are concentrated in close proximity to the red giant branch luminosity bump, the core-He burning stages, or the early-asymptotic giant branch. This is very clear from the \emph{Gaia}-based luminosities of the Li-rich giants. This is also seen when the CoRoT Li-rich giants are compared to a larger sample of 2252 giants observed in the CoRoT fields by the \emph{Gaia}-ESO Survey, which are distributed throughout the red giant branch in the $T_{\rm eff}$-$\log~g$ diagram. These observations show that the evolutionary stage is a major factor for the Li enrichment in giants. Other processes, such as planet accretion, contribute at a smaller scale.}


\maketitle

\section{Introduction}

Although more than three decades have past since the discovery of the first Li-rich giant \citep{1982ApJ...255..577W}, the origin of such objects remains without a clear explanation \citep[see][for a review]{2016Ap.....59..411L}. According to standard stellar evolution models, the surface Li abundances of low-mass red giants after the first dredge-up should lie below A(Li) $\sim$ 1.50 dex \citep[e.g.][]{2012A&A...543A.108L}. However, about 1-2\% of the known giants have been found to be richer in Li than this \citep[see, e.g.][]{2016MNRAS.461.3336C,2016ApJ...819..135K}.

Different scenarios have been proposed to explain their high-Li abundances. These scenarios can be broadly divided into those requiring internal fresh Li production, and those postulating external pollution by material rich in Li. A few additional processes have been proposed to explain some specific cases within the zoo of Li-rich giants, such as the red giant branch (RGB) phase transition discussed in \citet{2016A&A...585A.124C} and the extra-mixing inhibition discussed in \citet{2016A&A...591A..62S}.

\begin{table*}
 \caption[]{\label{tab:obs} Observational data of the new-Li rich giants.}
\centering\small
\begin{tabular}{cccccccr}
\hline
\hline
CNAME & Field & 2MASS ID & R.A. & DEC. & $V$ & RV & S/N \\
 & &  & h:m:s (J2000) & d:m:s (J2000) & mag & km s$^{-1}$ &  \\
\hline
08405643-5308309 & IC 2391 & 08405643-5308309 & 08:40:56.43 & $-$53:08:30.90 & 14.48 & +55.0 & 35 \\
17522490-2927512 & Rup 134 & 17522490-2927512 & 17:52:24.90 & $-$29:27:51.20 & 15.16$^{1}$ & +81.6 & 47 \\
17531013-2932063 & Rup 134 & 17531013-2932063 & 17:53:10.13 & $-$29:32:06.30 & 14.10$^{1}$ & $-$25.8 & 99 \\
18181062-3246291 & Bulge & 18181061-3246290 & 18:18:10.62 & $-$32:46:29.10 & 11.68$^{1}$ & +39.5 & 184 \\
18182698-3242584 & Bulge & 18182697-3242584 & 18:18:26.98 & $-$32:42:58.40 & 12.67$^{1}$ & +27.4 & 135 \\
18265248+0627259 & NGC 6633 & 18265248+0627259 & 18:26:52.48 & +06:27:25.90 & 14.45 & +32.7 & 51 \\
19223053+0138518 & Corot & 19223052+0138520 & 19:22:30.53 & +01:38:51.80 & 13.03 & $-$23.8 & 47 \\
19251759+0053140 & Corot & 19251759+0053141 & 19:25:17.59 & +00:53:14.00 & 14.55 & +88.7 & 39 \\
19261134+0051569 & Corot & 19261134+0051569 & 19:26:11.34 & +00:51:56.90 & 15.02 & +29.5 & 26 \\
19263808+0054441 & Corot & 19263807+0054441 & 19:26:38.08 & +00:54:44.10 & 13.28 & $-$57.8 & 66 \\
19264134+0137595 & Corot & 19264133+0137595 & 19:26:41.34 & +01:37:59.50 & 14.25 & +44.0 & 50 \\
19264917-0027469 & Corot & 19264917-0027469 & 19:26:49.17 & $-$00:27:46.90 & 15.83$^{2}$ & +77.9 & 20 \\
19265013+0149070 & Corot & 19265013+0149071 & 19:26:50.13 & +01:49:07.00 & 15.86 & $-$48.3 & 31 \\
19265193+0044004 & Corot & 19265195+0044004 & 19:26:51.93 & +00:44:00.40 & 13.09 & $-$12.9 & 116 \\
19270600+0134446 & Corot & 19270600+0134446 & 19:27:06.00 & +01:34:44.60 & 14.87 & +28.5 & 42 \\
19270815+0017461 & Corot & 19270815+0017461 & 19:27:08.15 & +00:17:46.10 & 15.26$^{1}$ & +31.3 & 24 \\
19273856+0024149 & Corot & 19273856+0024149 & 19:27:38.56 & +00:24:14.90 & 15.37 & +18.3 & 22 \\
19274706+0023447 & Corot & 19274706+0023448 & 19:27:47.05 & +00:23:44.70 & 14.78$^2$ & +46.5 & 27 \\
19280508+0100139 & Corot & 19280507+0100139 & 19:28:05.08 & +01:00:13.90 & 15.25 & +75.0 & 44 \\
19283226+0033072 & Corot & 19283226+0033072 & 19:28:32.26 & +00:33:07.20 & 14.77 & +49.8 & 32 \\
\hline
\end{tabular}
\tablefoot{The $V$ magnitudes are from APASS \citep{2015AAS...22533616H} unless otherwise noted: (1) The Guide Star Catalog, Version 2.3.2 (GSC2.3) (STScI, 2006), and (2) the NOMAD catalogue \citep{2004AAS...205.4815Z}. For the determination of radial velocities and values of signal-to-noise per pixel from the GIRAFFE spectra see \citet{2015A&A...580A..75J}. For the UVES spectra, the determination of these values is described in \citet{2014A&A...565A.113S}.}
\end{table*}

Internal Li production likely takes place by the mechanism proposed by \citet{1971ApJ...164..111C}. In this mechanism, the unstable isotope $^{7}$Be is produced in the stellar interior by the reaction $^{3}$He($\alpha$,$\gamma$)$^{7}$Be, which is followed by the decay $^{7}$Be($e^{-}$~$\nu$)$^{7}$Li, resulting in freshly synthesised $^{7}$Li. The Cameron \& Fowler mechanism, however, was introduced to explain observations of Li in asymptotic giant branch (AGB) stars. In AGB stars, the convective layer enters the H-burning
shell providing the means to bring the fresh Li to the surface \citep[see, e.g.][]{1975ApJ...196..805S,2011ApJ...741...26P}. In first-ascent red giants, however, the introduction of an extra mixing mechanism is required to bring the fresh Li to the surface before it is destroyed in a reaction with protons of the medium. The physical mechanism responsible for this fast deep mixing is still unknown \citep{1999ApJ...510..217S,2000A&A...359..563C,2001A&A...375L...9P}.

\begin{table*}
 \caption[]{\label{tab:par} Atmospheric parameters, lithium abundances, and rotational velocities for the newly discovered Li-rich giants.}
\centering\small
\begin{tabular}{ccccccccccccr}
\hline
\hline
CNAME & T$_{\rm eff}$ & $\sigma$ & $\log~g$ & $\sigma$ & [Fe/H] & $\sigma$ & $\xi$ & $\sigma$ & A(Li) & $\sigma$ & A(Li) & v$\sin~i$\\
            &   (K) & (K) &  &  &  &  & km s$^{-1}$ & km s$^{-1}$ & (LTE)  & & (non-LTE) & km s$^{-1}$\\
\hline
08405643-5308309 & 4486 & 142 & 2.54 & 0.15 & $-$0.12 & 0.12 &   -- &   -- & 2.64 & 0.18 & 2.60 & $\leq$7.0 \\
17522490-2927512 & 4644 & 188 & 2.80 & 0.14 &  +0.18 & 0.16 &   -- &   -- & 2.32 & 0.25 & 2.42 & $\leq$7.0\\
17531013-2932063 & 4557 & 169 & 2.72 & 0.18 &  +0.27 & 0.13 &   -- &   -- & 2.12 & 0.10 & 2.30 & $\leq$7.0 \\
18181062-3246291 & 4558 & 57 & 2.27 & 0.11 & $-$0.03 & 0.06 & 1.54 & 0.13 & 2.15 & 0.06 & 2.30 & 2.5$\pm$3.4 \\
18182698-3242584 & 4425 & 57 & 2.33 & 0.11 &  +0.10 & 0.13 & 1.53 & 0.29 & 4.12 & 0.06 & 4.04 & 6.0$\pm$4.6 \\
18265248+0627259 & 4982 & 192 & 2.88 & 0.23 & $-$0.08 & 0.23 &   -- &   -- & 2.74 & 0.10 & 2.69 & 37.1$\pm$2.0 \\
19223053+0138518 & 4579 & 42 & 2.49 & 0.09 &  +0.26 & 0.23 &   -- &   -- & 2.06 & 0.03 & 2.27 & $\leq$7.0\\
19251759+0053140 & 4621 & 38 & 2.78 & 0.09 &  +0.36 & 0.20 &   -- &   -- & 2.03 & 0.04 & 2.24 & $\leq$7.0 \\
19261134+0051569 & 4745 & 39 & 2.47 & 0.09 & $-$0.53 & 0.22 &   -- &   -- & 3.60 & -- & 3.25 & 7.5$\pm$1.7 \\
19263808+0054441 & 4655 & 31 & 2.82 & 0.09 &  +0.38 & 0.12 &   -- &   -- & 2.09 & 0.10 & 2.29 & 12.0$\pm$3.0 \\
19264134+0137595 & 4645 & 36 & 2.56 & 0.09 &  +0.28 & 0.20 &   -- &   -- & 3.60 & -- & 3.45 & $\leq$7.0 \\
19264917-0027469 & 4458 & 46 & 2.19 & 0.10 & $-$0.39 & 0.26 &   -- &   -- & 3.52 & 0.09 & 3.33 & 7.1$\pm$1.6 \\
19265013+0149070 & 4770 & 42 & 2.68 & 0.09 & $-$0.50 & 0.25 &   -- &   -- & 3.68 & 0.14 & 3.32 & 13.2$\pm$1.7 \\
19265193+0044004 & 4880 & 58 & 2.54 & 0.11 & $-$0.33 & 0.13 & 1.50 & 0.22 & 2.94 & 0.06 & 2.80 & 2.1$\pm$2.7 \\
19270600+0134446 & 4584 & 36 & 2.38 & 0.09 &  +0.19 & 0.17 &   -- &   -- & 3.67 & 0.13 & 3.53 & $\leq$7.0 \\
19270815+0017461 & 4514 & 41 & 2.28 & 0.09 & $-$0.29 & 0.26 &   -- &   -- & 2.33 & 0.08 & 2.37 & 7.8$\pm$1.6 \\
19273856+0024149 & 4446 & 38 & 2.39 & 0.09 & $-$0.16 & 0.21 &   -- &   -- & 2.33 & 0.16 & 2.37 & 12.0$\pm$0.5 \\
19274706+0023447 & 4608 & 66 & 3.21 & 0.13 &  +0.13 & 0.24 & -- & -- & 2.78 & 0.14 & 2.70 & 18.0$\pm$0.5 \\
19280508+0100139 & 4623 & 38 & 2.49 & 0.09 &  +0.11 & 0.18 &   -- &   -- & 3.66 & 0.10 & 3.49 & 8.2$\pm$1.5 \\
19283226+0033072 & 4600 & 38 & 2.64 & 0.09 &  +0.29 & 0.21 &   -- &   -- & 3.60 & -- & 3.46 & $\leq$7.0\\
\hline
\end{tabular}
\tablefoot{In the case of stars with GIRAFFE spectra, the errors in the parameters and abundances are the standard deviation of values obtained using multiple analysis pipelines. These are thus estimates of the internal error alone. The missing error of some Li abundances means that the values were determined by one single pipeline. Assuming a typical error of $\sim$ 0.10-0.15 dex (as for the other values) would be adequate in these cases. In the case of stars with UVES spectra, errors are obtained through modeling of how well the multiple pipelines reproduce the reference parameters of calibrating objects (e.g., \emph{Gaia} benchmark stars). The process of error estimation in the analysis of UVES spectra will be described in Casey et al. (in prep). For stars observed with GIRAFFE, values of v$\sin~i$ are reported only if above $\sim$ 7.0 km s$^{-1}$. For these measurements, the HR15N setup (centred at 665nm) was used.}
\end{table*}

The engulfment of planets and/or planetesimals is often the preferred scenario of external pollution to explain Li-rich giants \citep[see, e.g.][]{1999MNRAS.308.1133S,2016ApJ...829..127A,2016A&A...587A..66D,2016A&A...589A..57R}. The location in the Herztsprung-Russel (HR) diagram of Li-rich giants previously discovered within the \emph{Gaia}-ESO Survey seemed to be consistent with those of giants that engulfed close-in giant planets before evolving up the RGB \citep{2016MNRAS.461.3336C}. In this scenario, the Li enhancement should be accompanied by enhancement of other light elements, such as $^6$Li and Be. Enhancement in these elements, however, has never been detected in any of the Li-rich giants investigated so far \citep{1997A&A...321L..37D,1999A&A...345..249C,2005A&A...439..227M,2017PASJ...69...74T,2018arXiv180104379A}. Moreover, at least in red giants, the complexity of the evolutionary mixing events affecting Li and other elements has so far precluded the discovery of clear abundance signatures related to planet engulfment \citep[see, e.g.,][]{2016ApJ...818...25C,2016ApJ...827..129C}. This kind of joint Li and Be enhancement seems to have been detected in only one main-sequence star so far: in the open cluster NGC 6633, as reported by \citet{2005MNRAS.363L..81A}.

It has been suggested that the Li enhancement in giants might be connected to a phase of enhanced mass loss \citep{1996ApJ...456L.115D,2015ApJ...806...86D}. In this context, the star KIC\,4937011, located in the field of the open cluster NGC\,6819, is very interesting. The asteroseismic analysis of stars in NGC\,6819 by \citet{2011ApJ...739...13S} classified KIC\,4937011 as a non-member of the cluster. The oscillation pattern of KIC\,4937011 was found to be different from that of other stars at similar position in the colour-magnitude diagram (CMD). This star was later found to be Li-rich by \citet{2013ApJ...767L..19A}. They suggested that the asteroseismic mismatch could be related to the process that caused the Li enrichment. \citet{2015ApJ...802....7C} investigated KIC\,4937011
in detail and summarised all the evidence supporting cluster membership (such as radial velocity and overall chemical composition). More recently, \citet{2017MNRAS.472..979H} suggested that 
KIC\,4937011 is indeed a member of NGC\,6819, but that it experienced very high mass loss during its evolution (it has now 0.7~M$_{\odot}$ compared to 1.6~M$_{\odot}$ of other red giants in the cluster). 

Asteroseismology, thus, seems to offer the means to uncover new information about Li-rich giants. Four other Li-rich giants observed with Kepler \citep{2010Sci...327..977B} have been reported  \citep{2014ApJ...784L..16S,2015A&A...584L...3J,2018arXiv180410955B}. Nine Li-rich giants observed with CoRoT \citep[Convection, Rotation and planetary Transits,][]{2006cosp...36.3749B,2009A&A...506..411A} have been discovered with \emph{Gaia}-ESO data and were reported in \citet{2016MNRAS.461.3336C}. 

This work is the second \emph{Gaia}-ESO paper on the subject of Li-rich giants. Here, we report the discovery of 20 new Li-rich giants, 14 of which are in the CoRoT fields. These new Li-rich giants were discovered in new \emph{Gaia}-ESO observations that were not available in the data release used in \citet{2016MNRAS.461.3336C}. With the new stars, there are now 40 Li-rich giants identified using \emph{Gaia}-ESO data. Our goal is to discuss the properties of these new Li-rich giants and discuss which clues they provide about the origin of the Li enrichment. In particular, this offers the opportunity to revisit the conclusions of \citet{2016MNRAS.461.3336C} using the more recent reanalysis of \emph{Gaia}-ESO data available in the Survey's fifth internal data release (iDR5). We also take advantage of new PARSEC isochrones \citep{2018MNRAS.476..496F} and parallaxes of the recent second data release (DR2) of \emph{Gaia} \citep{2016A&A...595A...1G,2018arXiv180409365G,2018arXiv180409366L} to revisit the discussion about the position of the Li-rich giants in the HR diagram.

This paper is organised as follows. In Sect.\ \ref{sec:data} we give a brief description of the \emph{Gaia}-ESO Survey \citep{2012Msngr.147...25G,2013Msngr.154...47R} data and analysis. In Sect.\ \ref{sec:new} we present the properties of the 20 newly discovered Li-rich giants. In Sect.\ \ref{sec:disc} we use CoRoT data and \emph{Gaia}-based luminosities to update the discussion about the origin of the Li enrichment in the Li-rich giants discovered by the \emph{Gaia}-ESO Survey. Finally, Sect.\ \ref{sec:end} summarises our findings.

\section{Data and analysis}\label{sec:data}

The spectra used here have been obtained with the FLAMES \citep[Fibre Large Array Multi Element Spectrograph,][]{2002Msngr.110....1P} instrument at the European Southern Observatory's (ESO) Very Large Telescope (VLT) in Cerro Paranal, Chile. FLAMES was used to feed both the GIRAFFE medium-resolution (R $\sim$ 20\,000) and the UVES \citep[Ultraviolet and Visual Echelle Spectrograph,][]{2000SPIE.4008..534D} high-resolution (R $\sim$ 47\,000) spectrographs. 

Basic information on the newly discovered Li-rich giants is given in Table \ref{tab:obs}. All stars were observed with GIRAFFE, except for the two giants towards the bulge and the CoRoT target with CNAME~19265193+0044004, which were observed with UVES. The Li line at 6708\AA\ falls within the GIRAFFE setup HR15N ($\lambda\lambda$ 647-679 nm). In \emph{Gaia}-ESO, this GIRAFFE setup is only used to observe calibrators (such as the CoRoT stars) and stars in open clusters. All other stars in Milky Way fields are observed only with HR10 and/or HR21 and therefore the determination of Li abundances is not possible for them. UVES, on the other hand, is used to observe mainly FG-type dwarfs in the solar neighbourhood, giants in the field of open clusters, and towards the Bulge, or (bright) calibrators, such as some CoRoT giants. Because of these details of how the stars are observed, Li-rich giants have been discovered with \emph{Gaia}-ESO data only in these three types of fields (open clusters, CoRoT fields, or towards the Bulge).

\begin{figure}
\centering
\includegraphics[height = 7cm]{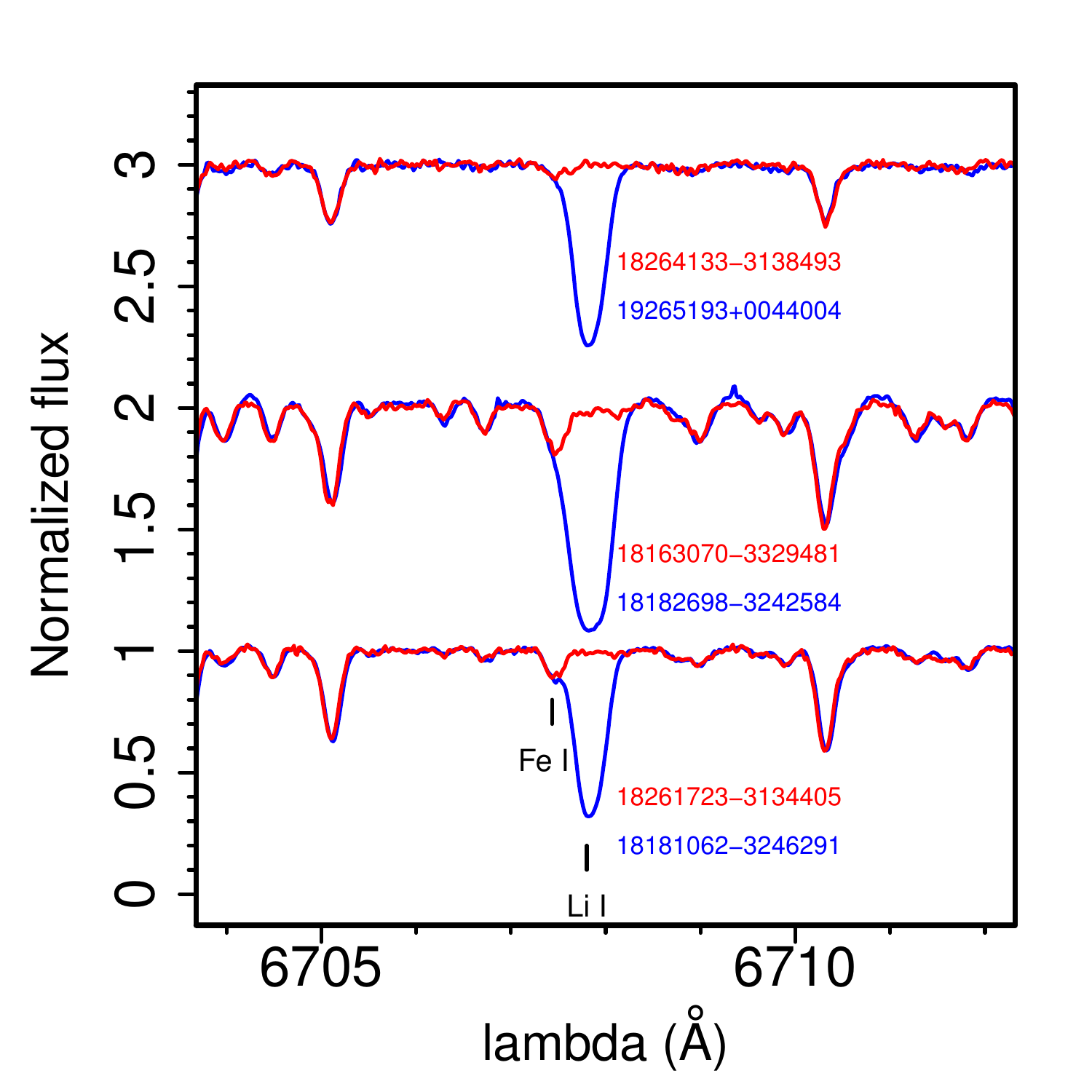}
\caption{Spectra of three Li-rich giants at the wavelength of the Li 6708 \AA\ line (in blue). In each case, the spectrum of a comparison giant with similar atmospheric parameters is also shown (in red; similar means within $\pm$ 50 K in $T_{\rm eff}$ and $\pm$ 0.10 dex in $\log~g$ and [Fe/H]). The flux has been normalised and arbitrarily shifted to facilitate the visualisation.}\label{fig:spec}
\end{figure}
\begin{figure*}
\centering
\includegraphics[height = 14cm]{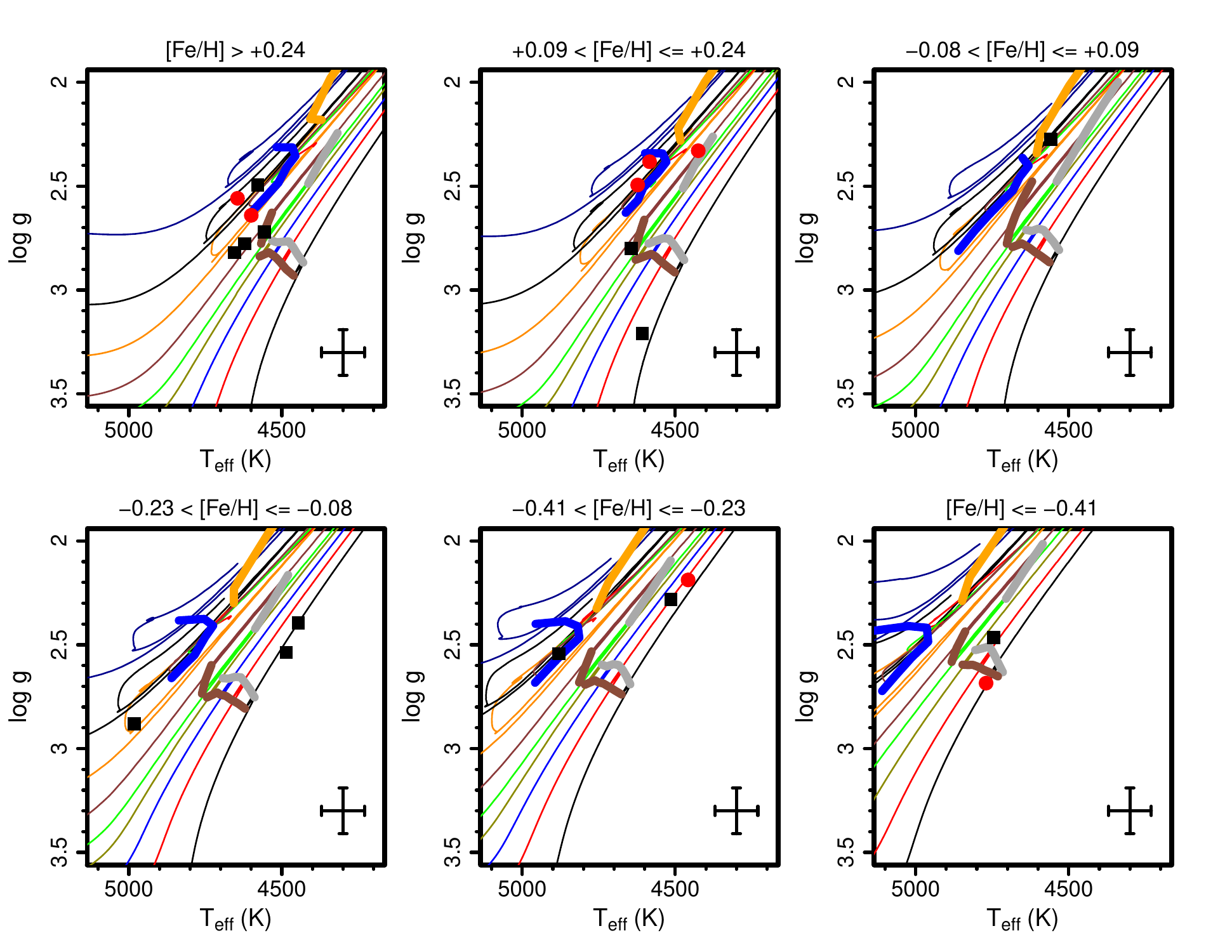}\caption{Newly discovered Li-rich giants in the $T_{\rm eff}$-$\log~g$ diagram divided according to metallicity into different panels. PARSEC evolutionary tracks \citep{2012MNRAS.427..127B,2018MNRAS.476..496F} of masses 0.8, 1.0, 1.2, 1.4, 1.5, 1.7, 2.0, 2.4, and 3.0 ~M$_{\odot}$ are shown. From the top left to the bottom right panels, the isochrones have [Fe/H] = +0.30, +0.18, 0.00, $-$0.15, $-$0.30,
and $-$0.52 dex. The range of [Fe/H] of the stars is given at the top of each panel. The beginning and the end of the RGB luminosity bump are marked as thick grey and brown lines, respectively. The position of the clump of low-mass giants is shown as a thick blue line. The beginning of the early-AGB of intermediate-mass stars is highlighted as the thick orange line. Super Li-rich giants with A(Li) $>$ 3.3 dex (in non-LTE) are shown as red circles, giants with Li abundance below this are shown as black squares. For the intermediate-mass stars, tracks are plotted only until the point where central He reaches approximately zero (end of core-He burning). Typical error bars ($\pm$ 70 K in $T_{\rm eff}$ and $\pm$ 0.10 dex in $\log~g$) are shown in the bottom right corner of the panels.}\label{fig:newgiants}
\end{figure*}

An overview of the GIRAFFE data reduction can be found in \citet{2015A&A...580A..75J}. The UVES data reduction is described in \citet{2014A&A...565A.113S}. 

The atmospheric parameters, effective temperature ($T_{\rm eff}$), surface gravity ($\log~g$), metallicity ([Fe/H]), and microturbulence ($\xi$), projected rotational velocities ($v~\sin~i$), and the chemical abundances are part of iDR5. The values of atmospheric parameters are given in Table \ref{tab:par}. The spectra were analysed using the \emph{Gaia}-ESO multiple pipelines strategy. For the case of stars in the field of young clusters, which can also contain pre-main sequence stars, the spectrum analysis was described in \citet{2015A&A...576A..80L}. The analysis of UVES spectra of other FGK-type stars was described in \citet{2014A&A...570A.122S}, and an updated discussion will be presented in Casey et al. (in prep.). The analysis of GIRAFFE spectra of other FGK-type stars was briefly described in \citet{2014A&A...567A...5R}. The results obtained in the different analyses are homogenised using the comprehensive set of stellar calibrators described in \citet{2017A&A...598A...5P}. The homogenisation process will be described in Hourihane et al. (in prep.). A brief description of the \emph{Gaia}-ESO atomic and molecular data is given in \citet{2015PhyS...90e4010H}. The analysis made use of the MARCS model atmospheres \citep{2008A&A...486..951G}.

The Li abundances were determined from the 6708~\AA\ line assuming local thermodynamical equilibrium (LTE). A combination of methods including spectrum synthesis, measurement of equivalent widths, and curve of growths was used. The line list includes the hyperfine structure and isotopic splitting of the Li line and the \ion{Fe}{i} blend at 6707.4 \AA. We selected from the iDR5 catalogue giants ($\log~g \leq 3.5$ dex) with $T_{\rm eff}$ between 4000 and 5000~K, and with detected Li abundances, not upper limits, where A(Li) $\geq$ 2.0 dex (in LTE). Example spectra of three Li-rich giants are shown in Fig.\ \ref{fig:spec}.

Originally, 21 candidate Li-rich giants were identified. Upon inspection of the spectra, however, one giant was found to lack an enhanced Li line (CNAME~08113284-4721004) and was excluded from further discussion. Its enhanced Li abundance is likely an artefact introduced by some failure of the analysis pipelines. We have checked the literature, and to the best of our knowledge, all 20 giants are reported to be Li-rich here for the first time.

Corrections for non-LTE effects were applied using the grid of \citet{2009A&A...503..541L}. The corrections depend on the atmospheric parameters and on the LTE Li abundance. They are usually positive if the Li enrichment is not too high (around 2.0~dex), but may become negative for the giants with Li abundances above $\sim$2.6~dex. For the stars observed with GIRAFFE, a spectroscopic determination of the microturbulence velocity was not possible. For the purposes of the non-LTE correction, a value of $\xi$ = 1.5~km s$^{-1}$ was adopted as is reasonable for red giants. In any case, the non-LTE correction is not very sensitive to $\xi$. The LTE and non-LTE Li abundances are given in Table \ref{tab:par}.

\section{Properties of the new Li-rich giants}\label{sec:new}

\begin{table*}
 \caption[]{\label{tab:pecul} Summary of the noteworthy characteristics of the new Li-rich giants as discussed in Sections \ref{sec:new} and \ref{sec:disc}.}
\centering\small
\begin{tabular}{cl}
\hline
\hline
CNAME & Characteristics \\
  &  \\
\hline
08405643-5308309 &  Above the bump in both the $T_{\rm eff}$-$\log~g$ diagram (Fig. \ref{fig:newgiants}) and the HR-diagram (Fig. \ref{fig:lum}); bottom left panels \\
17522490-2927512 &  \\
17531013-2932063 &  [$\alpha$/Fe]  $>$ +0.15 \\
18181062-3246291 &  \\
18182698-3242584 &  $\alpha$-enhanced with super-solar metallicity \\
18265248+0627259 &  Possible core-He burning intermediate-mass giant? (bottom left panel of Fig. \ref{fig:newgiants}); candidate for \\ 
                                 & planet engulfment (fast rotation); closer to the clump in the HR-diagram (Fig. \ref{fig:lum}) \\
19223053+0138518 &  \\
19251759+0053140 &  \\
19261134+0051569 &  S/N $<$ 30; [$\alpha$/Fe]  $>$ +0.15  \\
19263808+0054441 &  \\
19264134+0137595 &  \\
19264917-0027469 &  Above the bump ($T_{\rm eff}$-$\log~g$ diagram, bottom-middle panel of Fig. \ref{fig:newgiants}); S/N $<$ 30; [$\alpha$/Fe]  $>$ +0.15;  \\
                                 & luminosity not computed \\
19265013+0149070 &  [$\alpha$/Fe]  $>$ +0.15; Ba enhanced (mass transfer from an AGB?) \\
19265193+0044004 &  Core-He burning giant from seismic data\\
19270600+0134446 &  Infrared excess at 22$\mu$m \\
19270815+0017461 &  Above the bump ($T_{\rm eff}$-$\log~g$ diagram, bottom left panel of Fig. \ref{fig:newgiants}); S/N $<$ 30; [$\alpha$/Fe]  $>$ +0.15;  \\
                                 &  luminosity not computed \\
19273856+0024149 &  Above the bump ($T_{\rm eff}$-$\log~g$ diagram, bottom left panel of Fig. \ref{fig:newgiants}); S/N $<$ 30; [$\alpha$/Fe]  $>$ +0.15; \\
                                  & luminosity  not computed \\
19274706+0023447 &  Below the bump ($T_{\rm eff}$-$\log~g$ diagram, top middle panel of Fig. \ref{fig:newgiants}); S/N $<$ 30; candidate for planet \\
                                 & engulfment (fast rotation); at the clump in the HR-diagram (top middle panel of Fig. \ref{fig:lum}) \\
19280508+0100139 &  \\
19283226+0033072 &  \\
\hline
\end{tabular}
%\tablefoot{}
\end{table*}

Within the \emph{Gaia}-ESO data analysis cycle, a detailed system of flags has been defined with which any peculiarity identified in the spectra and/or issues found during the analysis can be
reported. It is anticipated that a number of these flags will be included in later public data releases (flags to be described in Van Eck et al., in prep). For the moment, we can report that no important problem with the spectra of these Li-rich giants has been flagged (e.g. no emission lines, and no evidence of multiplicity in the spectra). For the method developed to identify spectroscopic binaries, we refer to the discussion presented in \citet{2017A&A...608A..95M}.

The subsections below report our investigation of the properties of the new Li-rich giants (e.g. core of strong lines for signs of magnetic activity, infrared (IR) magnitudes, rotational velocities, chemical abundances, and Galactic velocities). We attempted to identify characteristics in common among these stars. A summary of these findings is given in Table \ref{tab:pecul}.

\begin{table*}
 \caption[]{\label{tab:mass} Masses, ages, and distances estimated using {\sf UniDAM}.}
\centering\small
\begin{tabular}{ccccc}
\hline
\hline
CNAME & Mass & $\log({\rm Age})$ & Distance & Stage \\
  & (M$_{\odot}$) & (Gyr) & (kpc)  & \\
\hline
08405643-5308309 & 1.2$\pm$0.3  & 9.7$\pm$0.3 & 2.5-2.7$\pm$0.5 & I or II \\
17522490-2927512 & -- & -- & -- & --  \\
17531013-2932063 & -- & -- & -- & --  \\
18181062-3246291 & 1.2-2.1$\pm$0.5 & 9.1-9.8$\pm$0.4   & 1.7-2.6$\pm$0.5   & I, II or III\\
18182698-3242584 & -- & -- & -- & --   \\
18265248+0627259 & 1.2-1.9$\pm$0.5  & 9.3-9.8$\pm$0.3 & 2.5-3.8$\pm$0.7 & I or II \\
19223053+0138518 & 1.4$\pm$0.4  & 9.6$\pm$0.3 & 1.5-1.6$\pm$0.2 & I or II \\
19251759+0053140 & 1.4-1.9$\pm$0.2 & 9.2-9.6$\pm$0.2 & 1.9-2.6$\pm$0.3 & I or II \\
19261134+0051569 & 1.2$\pm$0.3 & 9.7$\pm$0.3 &  4.0$\pm$0.6 & I or II \\
19263808+0054441 & 1.4-1.9$\pm$0.2 & 9.2-9.6$\pm$0.2 & 1.0-1.4$\pm$0.2 & I or II \\
19264134+0137595 & 1.6-2.6$\pm$0.3  & 8.9-9.5$\pm$0.2  & 2.9-3.8$\pm$0.5 & II \\
19264917-0027469 & 1.3-1.9$\pm$0.3  & 9.2-9.7$\pm$0.3 & 8.5-11.6$\pm$0.1 & I or III \\
19265013+0149070 & 1.2-1.7$\pm$0.3  & 9.4-9.7$\pm$0.3 & 4.2-5.4$\pm$0.7 & I or II \\
19265193+0044004 & 1.3-1.4$\pm$0.4 & 9.5-9.6$\pm$0.3 & 1.8-1.9$\pm$0.3 & I or II  \\
19270600+0134446 & 1.1-1.6$\pm$0.4  & 9.5-9.8$\pm$0.3 & 3.5-3.9$\pm$0.8 & I or II \\
19270815+0017461 & 1.2-1.3$\pm$0.4  & 9.7-9.8$\pm$0.3 & 5.1-5.5$\pm$1.0 & I or II \\
19273856+0024149 & 1.1-1.2$\pm$0.2  & 9.8-9.9$\pm$0.2 & 4.3-4.5$\pm$0.6 & I or II \\
19274706+0023447 &  -- & -- & -- & --  \\
19280508+0100139 & 1.3-1.4$\pm$0.4 & 9.6-9.7$\pm$0.3 & 4.6-4.9$\pm$0.7 & I or II \\
19283226+0033072 & 1.3$\pm$0.2  & 9.7$\pm$0.2 &  2.7$\pm$0.4 & I \\
\hline
\end{tabular}
\tablefoot{We adopt the mean values of mass, age, and distance and the standard deviation, but note that other estimates are also provided by {\sf UniDAM} (e.g. mode and median). Usually, {\sf UniDAM} returns two (or three) estimates per star, assuming different evolutionary stages -- I: pre He-core burning; II: during He-core burning, or III: post He-core burning. The stages are given in the last column. When the estimates per stage are different, we list all values, otherwise only one value is given. Solutions of low quality (low probability) are discarded.}
\end{table*}

\subsection{Position in the $T_{\rm eff}$-$\log~g$ diagram}

Fig.~\ref{fig:newgiants} displays the newly discovered Li-rich giants in $T_{\rm eff}$-$\log~g$ diagrams, where the giants are divided according to metallicity. A selection of evolutionary tracks for masses between 0.8 and 3.0~M$_{\odot}$ is shown for comparison. The regions of the RGB luminosity bump, the clump of low-mass stars, and of the beginning of the early-AGB are indicated. Low-mass stars here are those stars that go through the He-core flash at the end of the RGB. The beginning of the early-AGB is defined here as the first point where the luminosity produced by the He-burning shell becomes higher than the luminosity produced by the H-burning shell, at the evolutionary stages after the abundance of central He has reached zero.

For masses below 1.4 M$_{\odot}$, we use the new PARSEC evolutionary tracks of \citet{2018MNRAS.476..496F} which were computed with a new envelope overshooting calibration. As a result, the RGB bump, for these tracks, is shifted by between +0.15 or +0.20 dex in $\log~g$ in comparison with the older PARSEC tracks. For masses above 1.4 M$_{\odot}$, we plot the older PARSEC tracks (as the position of the RGB bump was not changed in the new tracks).

 \citet{2000A&A...359..563C} suggested that Li-rich giants were preferentially located in two regions of the HR diagram. Low-mass Li-rich giants would be located at the RGB bump, and intermediate-mass Li-rich giants would be located at the early AGB. \citet{2000A&A...359..563C} connected the Li enrichment with an extra-mixing process that activates at these evolutionary stages.

More recently, some Li-rich giants were found to be instead core He-burning giants \citep{2011ApJ...730L..12K,2014A&A...564L...6M,2014ApJ...784L..16S,2018arXiv180410955B}. There are, however, works that report Li-rich giants throughout the RGB \citep[see, e.g.][]{2011A&A...531L..12A,2011A&A...529A..90M,2013MNRAS.430..611M} and others that find Li-rich objects also among less evolved stars \citep{2011ApJ...738L..29K,2016A&A...589A..61G,2018ApJ...852L..31L}. 

Fig. \ref{fig:newgiants} shows at least one giant (18265248+0627259) close to the position of the core He-burning stage at the intermediate-mass regime (in the bottom left panel). This star is potentially very interesting. To explain a concentration of Li-rich giants around the clump, \citet{2011ApJ...730L..12K} suggested an episode of Li production related to the He-core flash. However, such intermediate-mass giants do not go through the He-core flash and would thus require a different scenario for the Li enrichment. Nevertheless, this star is a fast rotator (see Section \ref{sec:rot} below) and has quite uncertain parameters. Thus, it might also be a low-mass clump giant.

Within the errors, the position of most giants in Fig. \ref{fig:newgiants} is consistent with the RGB luminosity bump or the clump of low-mass stars. This is true in all metallicity intervals. There are maybe five giants (one in the top middle panel, two in the bottom left panel, and two in the bottom middle panel) that fall either above or below the bump. For at least four of them, we consider the parameters to be uncertain. These four stars out of the five have spectra with signal-to-noise (S/N) below 30 (see Table \ref{tab:obs}). They might have been excluded from the discussion, but we chose to report them here, nonetheless, because they have enhanced Li lines that make them genuine Li-rich giants.

Given the error bars, a few of the giants might instead be in the early-AGB region, if they are of intermediate mass. The error bars do not allow us to clearly classify them as low- or intermediate-mass giants. Despite the difficulty in assigning a specific evolutionary stage, it seems clear from Fig. \ref{fig:newgiants} that the Li-rich giants are found in a narrow and specific region of the diagram.

In general, the position of giants in such spectroscopic diagrams can be quite uncertain. A plot such as we show in Fig.\ \ref{fig:newgiants} is not sufficient to tell the evolutionary stages apart. Photometric diagrams tend to be more precise if the distance to the star and the reddening are well known. We resume this discussion using \emph{Gaia} DR2 parallaxes in Section \ref{sec:gaia}. Otherwise, asteroseismology is the only way to properly separate the evolutionary stages \citep{2011Natur.471..608B,2011A&A...532A..86M,2017MNRAS.466.3344E}. The stellar properties based on CoRoT data, which are available for some of our giants, are discussed in Sections \ref{sec:seismic} and \ref{sec:corot}.

\begin{figure*}
\centering
\includegraphics[height = 7cm]{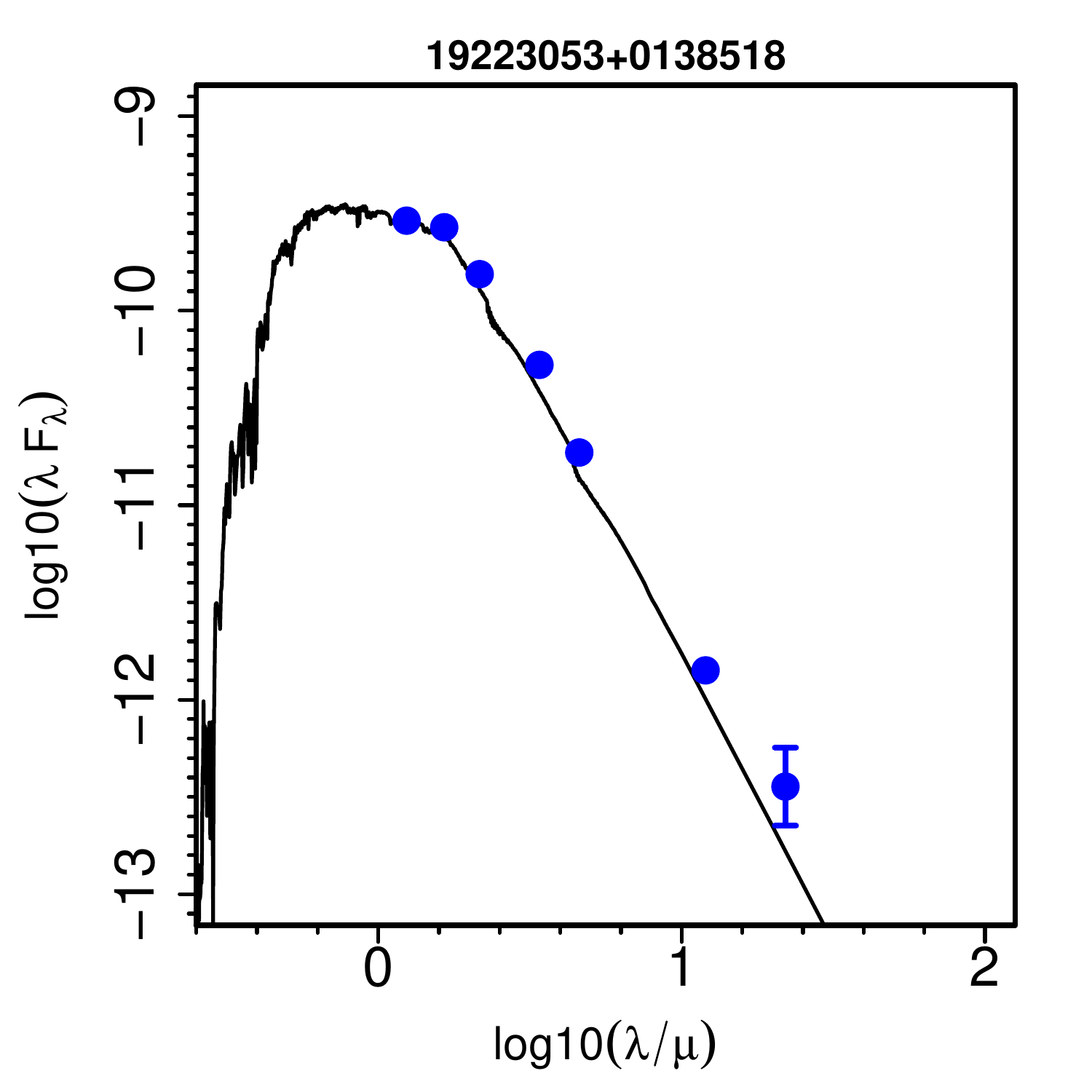}
\includegraphics[height = 7cm]{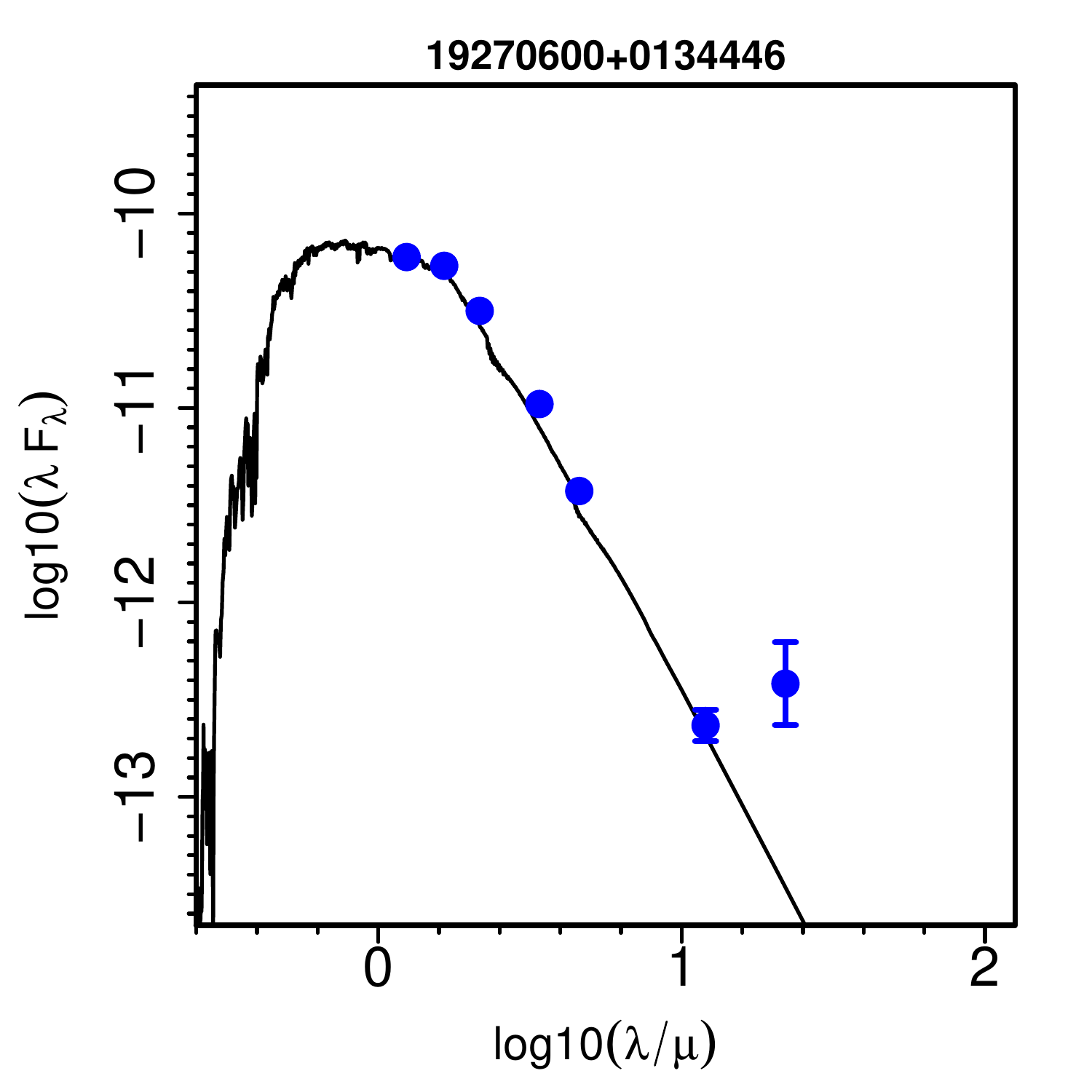}
\caption{Comparison between synthetic flux model and magnitudes (2MASS and WISE) for two stars, 19223053+0138518 (left) and 19270600+0134446 (right). Star 19270600+0134446 seems to have excess emission at the $W4$ band (22 $\mu$m).}\label{fig:ir}
\end{figure*}

\subsection{Stellar masses}\label{sec:unidam}

\citet{2017A&A...604A.108M} recently presented a tool for estimating masses, ages, and distances of stars using a Bayesian approach ({\sf UniDAM}, the unified tool to estimate distances, ages, and masses). The tool makes use of PARSEC isochrones \citep{2012MNRAS.427..127B} and needs as input the atmospheric parameters ($T_{\rm eff}$, $\log~g$, and [Fe/H]) and the IR photometry from 2MASS \citep[Two Micron All-Sky Survey,][]{2006AJ....131.1163S} and WISE \citep[Wide-field Infrared Survey Explorer,][]{2010AJ....140.1868W,2012yCat.2311....0C,2013yCat.2328....0C}.

Using {\sf UniDAM}, \citet{2017A&A...604A.108M} estimated the mass of one star of our sample, CNAME 19223053+0138518, using \emph{Gaia}-ESO DR2 atmospheric parameters. They derived mean masses of 1.31 or 1.50 ($\pm$ 0.21) M$_{\odot}$, depending on whether the star was assumed to be before or during He-core burning.

We have employed {\sf UniDAM} to obtain an indicative value of the stellar masses in our sample. We used the atmospheric parameters from Table \ref{tab:par} combined with $J$, $H$, and $Ks$ magnitudes from 2MASS, and $W1$ and $W2$ magnitudes from WISE. All magnitudes were retrieved from the Vizier database \citep{2000A&AS..143...23O}. The results are given in Table \ref{tab:mass} (and also include age and distance estimates). For a few stars the solutions did not converge or were flagged as being of low quality, and are thus not given. Estimates are not provided for star CNAME 17531013-2932063 as it has saturated 2MASS magnitudes. 
Some discrepancy between the masses estimated using {\sf UniDAM} and what would be estimated by eye from Fig. \ref{fig:newgiants}
can be noted. This is likely caused by the additional use of IR photometry in the calculations. Within the errors, however, the estimates of stellar masses would be in agreement. Moreover, we tested {\sf UniDAM} in a series of giants in open clusters observed by \emph{Gaia}-ESO with ages between 0.3 and 4.5 Gyr. The cluster ages and stellar masses obtained by {\sf UniDAM} were in agreement, within the errors, with the known properties of the clusters.

The {\sf UniDAM} code typically outputs two (sometimes three) solutions, assuming that the giants are in different evolutionary stage. Most Li-rich giants seem to be low-mass stars between 1.1 and 1.4 M$_{\odot}$, either before or during core-He burning, and thus either at the RGB luminosity bump or at the clump.

We remark that this method does not use any seismic information, but only spectroscopic and photometric observables and theoretical isochrones. It is well known that such mass estimates for red giants are affected by large uncertainties, given the accumulation of model tracks with different masses in a small region of the HR diagram and uncertainties in the chemical composition. In Section \ref{sec:seismic} below, we also discuss estimates of stellar masses based on seismic properties. Although these values also suffer from large uncertainties, the estimates seem to be consistent within the errors.

\subsection{Activity, infrared excess, and rotation}\label{sec:rot}

Lithium-rich giants are sometimes found as by-products in searches for young stars that use x-ray detection, IR excess, or chromospheric activity as selection criteria \citep[e.g.][]{1993A&A...268L..25G,1998A&AS..127..139C,2018arXiv180100671F}. Several works have reported detection of magnetic activity in some Li-rich giants \citep[see, e.g.][]{2013A&A...551A...2K,2017A&A...606A..42K,2014A&A...571A..74K}. To search for evidence of activity, we investigated the stellar spectra for signs of emission in the core of strong lines (H$\alpha$ and the near-IR \ion{Ca}{ii} lines). No clear signs of emission were found in any of the giants.

The IR excess that is sometimes reported in Li-rich giants has been suggested to be connected to an episode of enhanced mass loss \citep{1996ApJ...456L.115D,1997ApJ...482L..77D,2015ApJ...806...86D}. However, investigation of large samples of Li-rich and Li-normal giants has shown that IR excess seems to be rare \citep{1999A&A...342..831J,2015A&A...577A..10B,2015AJ....150..123R}. This indicates that either the possible mass-loss event is very short lived or that there is no connection between mass loss and Li enrichment.

\begin{table*}
 \caption[]{\label{tab:abun} Selected chemical abundances of the Li-rich giants.}
\centering\small
\begin{tabular}{ccccccc}
\hline
\hline
CNAME & [$\alpha$/Fe] & [Mg/Fe] & [Al/Fe] & [Na/Fe] & [Ba/Fe] & [Eu/Fe] \\
\hline
08405643-5308309 & -- & -- &  -- & -- & -- & -- \\
17522490-2927512 & $-$0.17$\pm$0.17 & -- &  +0.21$\pm$0.07 & -- & -- & -- \\
17531013-2932063 & +0.18$\pm$0.13 & -- &  +0.13$\pm$0.06 & -- & -- & -- \\
18181062-3246291 & $-$0.01$\pm$0.08 & $-$0.02$\pm$0.05 & $-$0.04$\pm$0.03 & $-$0.07$\pm$0.09 & +0.02$\pm$0.13 & $-$0.14$\pm$0.06 \\
18182698-3242584 & +0.22$\pm$0.16 & +0.22$\pm$0.12 & +0.32$\pm$0.08 & +0.37$\pm$0.29 & +0.01$\pm$0.01 & $-$0.04$\pm$0.11\\
18265248+0627259 &  -- & -- & -- & -- & -- & -- \\
19223053+0138518 & +0.05$\pm$0.23 & +0.00$\pm$0.04 & +0.26$\pm$0.05 & -- & -- & -- \\
19251759+0053140 & $-$0.12$\pm$0.20 & $-$0.09$\pm$0.03 & +0.07$\pm$0.05 & -- & -- & -- \\
19261134+0051569 & +0.39$\pm$0.22 & +0.37$\pm$0.04 & +0.34$\pm$0.05 & -- & +0.23$\pm$0.21 & --\\
19263808+0054441 & $-$0.09$\pm$0.12 & +0.01$\pm$0.02 & +0.11$\pm$0.03 & -- & -- & -- \\
19264134+0137595 & +0.06$\pm$0.20 & +0.07$\pm$0.03 & +0.18$\pm$0.04 & -- & -- & -- \\
19264917-0027469 & +0.40$\pm$0.26 & +0.34$\pm$0.04 & +0.52$\pm$0.10 & -- & $-$0.20$\pm$0.24 & --\\
19265013+0149070 & +0.36$\pm$0.25 & +0.30$\pm$0.04 & +0.34$\pm$0.05 & -- & +0.77$\pm$0.24 & -- \\
19265193+0044004 & +0.05$\pm$0.13 & +0.04$\pm$0.08 & +0.05$\pm$0.05 & 0.00$\pm$0.13 & $-$0.10$\pm$0.03 & $-$0.06$\pm$0.02\\
19270600+0134446 & $-$0.03$\pm$0.17 & $-$0.05$\pm$0.03 & +0.22$\pm$0.05 & -- & -- & -- \\
19270815+0017461 & +0.16$\pm$0.26 & +0.16$\pm$0.04 & +0.29$\pm$0.05 & -- & +0.13$\pm$0.22 & -- \\
19273856+0024149 & +0.19$\pm$0.22 & +0.12$\pm$0.04 & +0.17$\pm$0.05 & -- & -- & +0.11$\pm$0.06 \\
19274706+0023447 & +0.07$\pm$0.24 & $-$0.03$\pm$0.03 & +0.12$\pm$0.05 & -- & -- & +0.27$\pm$0.06 \\
19280508+0100139 & +0.05$\pm$0.18 & $-$0.01$\pm$0.03 & +0.12$\pm$0.05 & -- & -- & +0.01$\pm$0.06 \\
19283226+0033072 & $-$0.06$\pm$0.22 & $-$0.05$\pm$0.03 & +0.13$\pm$0.05 & -- & -- & -- \\
\hline
\end{tabular}
\tablefoot{Solar abundances of Mg, Al, Na, Ba, and Eu were adopted from \citet{2007SSRv..130..105G}. The abundance errors have the same meaning as discussed in the note of Table \ref{tab:par}.}
\end{table*}

We have investigated the IR behaviour of our Li-rich giants using 2MASS and WISE photometry. For comparison, a flux model of each star was computed using Kurucz codes \citep{1993KurCD..13.....K}. The modelled log(wavelength$\times$flux) was normalised to the $J$ band and compared to the remaining magnitudes. The agreement between models and observations is very good for all bands from $H$ to $W3$. For most stars, the WISE $W4$ band (at 22 $\mu$m) is only an upper limit. Only two stars have $W4$ detections. For star CNAME 19223053+0138518 (CoRoT 100440565), the agreement with the model is good. For star CNAME 19270600+0134446 (CoRoT 101205220), an excess emission is indicated at 22 $\mu$m (Fig. \ref{fig:ir}). Given the lack of $W4$ magnitudes for most stars in our sample, we do not have a clear picture of how common the IR excess is in these new Li-rich giants.

Projected rotational velocities are listed in Table \ref{tab:par}. Many of the giants have only upper limits determined from their GIRAFFE spectra and are thus likely slow rotators, including 19270600+0134446, the only giant with IR excess. Nine giants, however, have v~$\sin~i$ $>$ 7-8 km s$^{-1}$. This seems to agree with the results of \citet{2002AJ....123.2703D}, who found that Li-rich giants are more common among fast-rotating giants (defined by them as giants with v~$\sin~i$ $>$ 8 km s$^{-1}$).

Fast rotation is one of the expected outcomes of planet engulfment \citep[e.g.][]{2009ApJ...700..832C}. The high v~$\sin~i$ of some of the Li-rich giants could thus be interpreted as a sign of engulfment. \citet{2016A&A...591A..45P,2016A&A...593A.128P} computed models that take into account the interaction between the planetary orbit and rotation in stars during engulfment episodes. In their rotating stellar models, equatorial velocities (not projected) of the order of 5-10 km s$^{-1}$ are possible for giants of 1.5 M$_{\odot}$ and $\log~g$ $\sim$ 2.5 dex, even without engulfment, just from the normal spin-down evolution of the star. Only higher v~$\sin~i$ values would need to be explained with some sort of acceleration of the stellar surface.

Two stars in our sample could be examples of such cases. One is star 19274706+0023447 (CoRoT 101314825), the giant at the lower RGB (top middle panel of Fig. \ref{fig:newgiants}), which has 18 km s$^{-1}$. The other is star 18265248+0627259, the fastest rotator in this sample with 37 km s$^{-1}$ (the star apparently at the core-He burning phase, bottom left panel of Fig. \ref{fig:newgiants}). In light of the work of \citet{2016A&A...591A..45P,2016A&A...593A.128P}, both stars might have suffered surface acceleration because of the engulfment of planets. We note that this last star was observed in the field of the open cluster NGC 6633 (with ID NGC 6633 JEF 49), but it is not a member based on its radial velocity (RV) and photometry \citep[][who also measured v~$\sin~i$ = 39 km $s^{-1}$]{1997MNRAS.292..177J}.

\subsection{Chemical abundances}

We have checked the chemical abundances of other elements available in iDR5 for possible anomalies. Abundances of a few selected elements in the Li-rich giants are given in Table \ref{tab:abun}. The abundance information for the stars observed with GIRAFFE is limited because of the restricted wavelength range of the spectra \citep[see][]{2014A&A...572A..33M}. Abundances of C, N, and O are only available for giants observed with UVES and are given in Table \ref{tab:CNO}.

Surface abundances of C and N are expected to be altered in giants that have gone through the first dredge-up; [C/Fe] $\sim$ $-$0.20 dex and [N/Fe] $\sim$ +0.40 dex \citep[see, e.g.][for some recent references]{2015A&A...573A..55T,2016A&A...595A..16T,2016A&A...589A..50D,2016MNRAS.462..794D,2015MNRAS.446.3562B,2016MNRAS.463..580B,2018MNRAS.474.4810S}. The C and N abundances seem consistent with the stars having experienced first dredge-up, although N in the star CNAME 18181062-3246291 seems to have been only mildly affected.

\begin{table}
 \caption[]{\label{tab:CNO} Abundances of C, N, and O in the three Li-rich giants observed with UVES. }
\centering\small
\begin{tabular}{cccc}
\hline
\hline
CNAME & [C/Fe] & [N/Fe] & [O/Fe]  \\
\hline
18181062-3246291 & $-$0.17$\pm$0.03 & +0.01$\pm$0.06 & $-$0.02$\pm$0.05 \\
18182698-3242584 & $-$0.03$\pm$0.04 & +0.46$\pm$0.06 & +0.01$\pm$0.05 \\
19265193+0044004 & $-$0.21$\pm$0.06 & +0.23$\pm$0.05 & $-$0.10$\pm$0.03 \\
\hline
\end{tabular}
\tablefoot{Solar abundances of C, N, and O were adopted from \citet{1998SSRv...85..161G}. We gave preference to these older values for CNO because \citet{2007SSRv..130..105G} list abundances derived using 3D models, which considerably decrease the value of the reference solar abundances. Our analysis, however, is based on 1D models, and thus \citet{1998SSRv...85..161G} offer more consistent reference values. All abundances are given in LTE. The abundance errors have the same meaning as discussed in the note of Table \ref{tab:par}.}
\end{table}

\begin{figure*}
\centering
\includegraphics[height = 6.5cm]{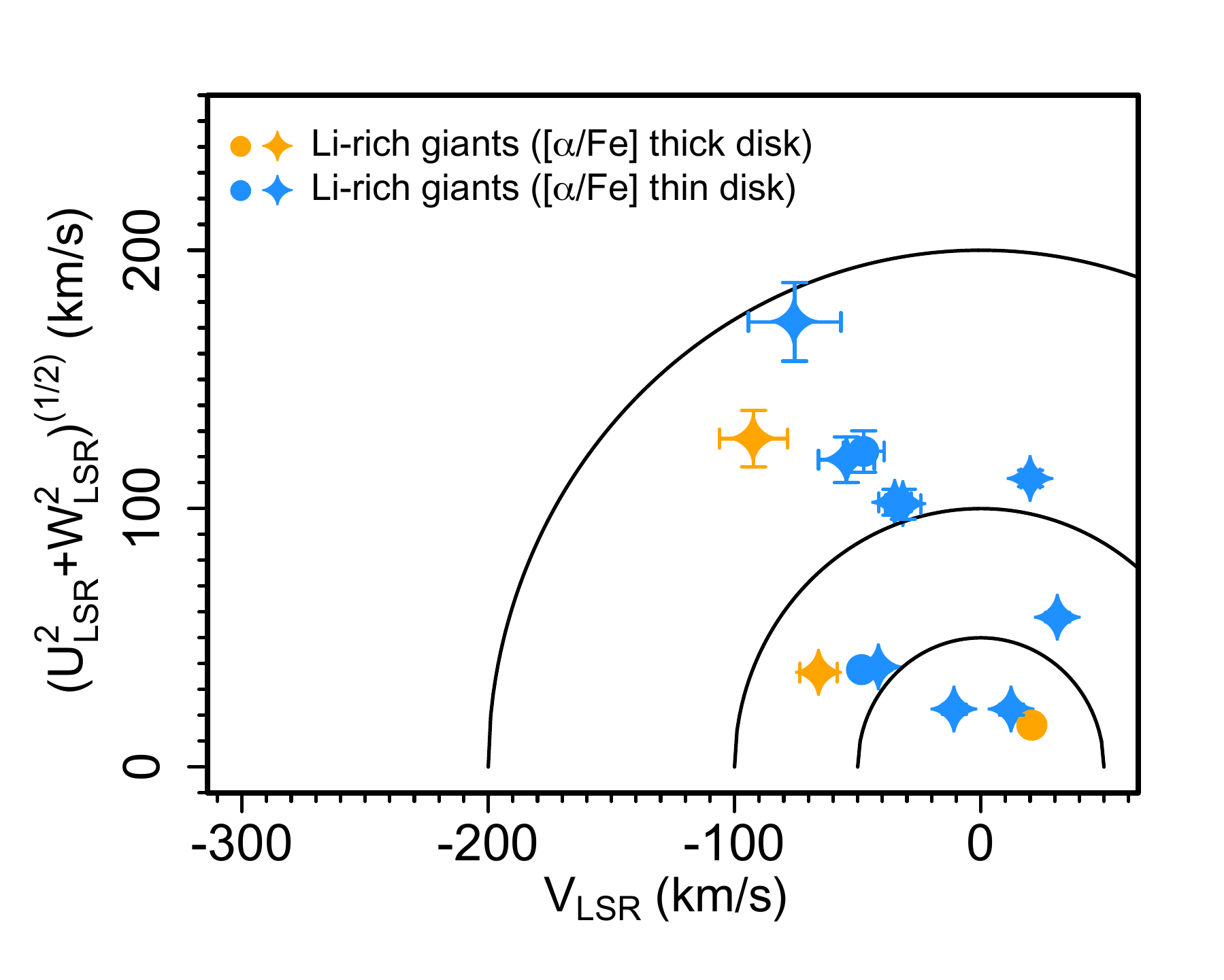}
\includegraphics[height = 6.5cm]{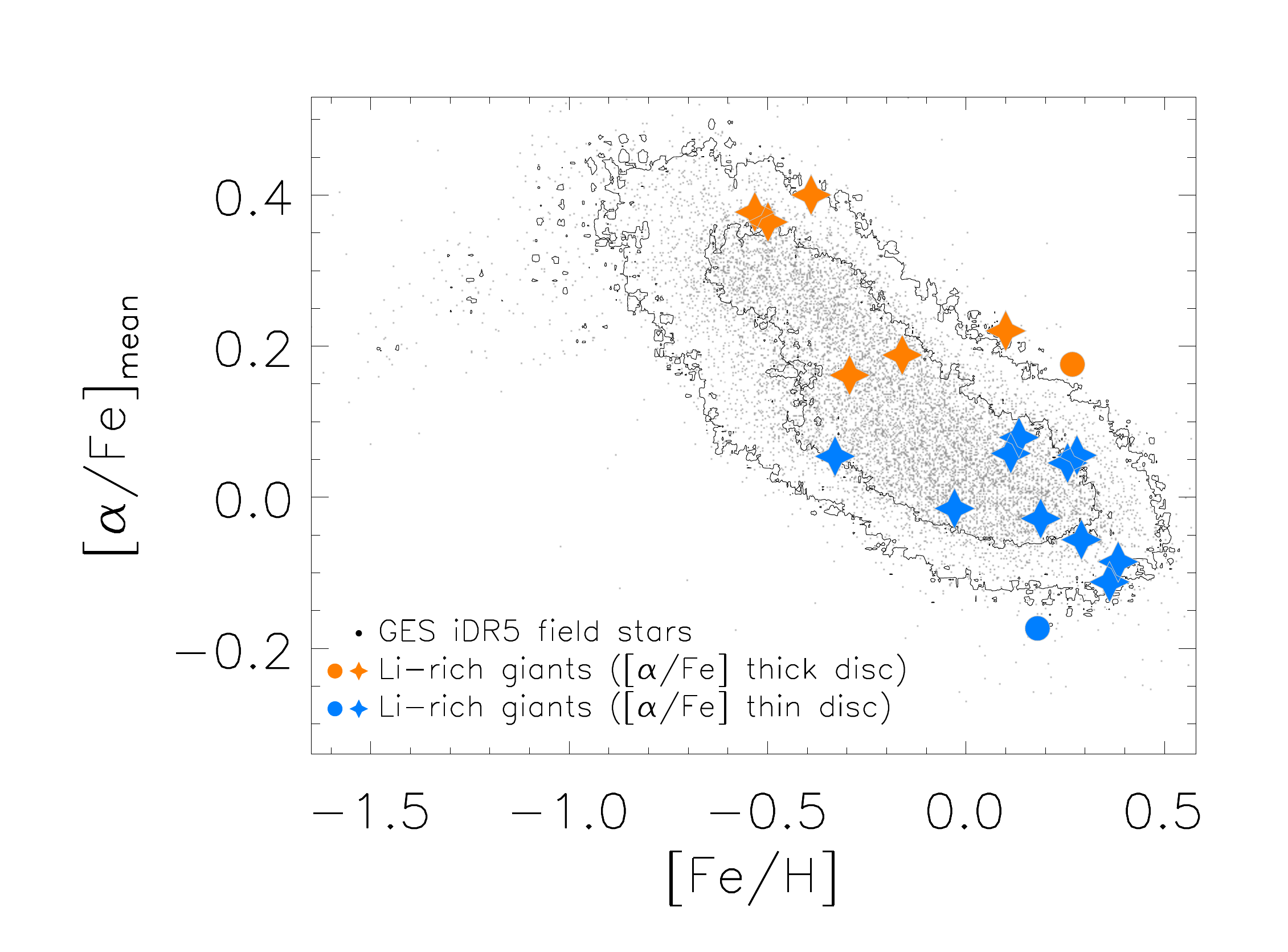}
\caption{\emph{Left}: Toomre diagram of the new Li-rich giants with good \emph{Gaia} DR2 parallaxes (i.e. only the 14 giants with a relative error of the parallax lower than 15\%). Error bars are shown, but in many cases they are smaller than the point size. \emph{Right:} Chemical plot of the [$\alpha$/Fe] ratio as a function of [Fe/H]. In all plots, the field Li-rich giants are displayed as stars and giants in fields of open clusters as solid circles. The colour is orange or blue for stars that
are tentatively associated  with the thick or thin disc, respectively. Other Gaia-ESO iDR5 field stars are shown as dots in the right panel}.\label{fig:grv}
\end{figure*}

Abundances of other elements (Na, Al, $\alpha$, iron peak, and neutron capture) were also investigated. Of the heavy neutron-capture elements, Table \ref{tab:abun} only lists [Ba/Fe] and [Eu/Fe] values. The only peculiarity that was identified lies in the [Ba/Fe] ratio of star CNAME 19265013+0149070 (CoRoT 101162874), which is clearly enhanced. We have double-checked the Ba abundance of this star using spectrum synthesis and confirm the reported enhancement. We remark that this star is rotating moderately fast (v~$\sin~i$ = 13.2 km s$^{-1}$, see Table \ref{tab:par}). The Ba enhancement might suggest that this could be a barium giant \citep[see, e.g.][]{2016MNRAS.459.4299D,2017A&A...608A.100E}. Barium giants are members of binary systems, with companions that are currently white dwarfs. The more massive companion evolved faster, went through the AGB phase and enriched itself with neutron capture elements, before transferring mass to the current Ba giant. The Li enhancement might also originate from the AGB companion. However, there seems to be no clear evidence of other Li-rich Ba giants \citep{1993PASP..105..568L}.

\begin{table}
 \caption[]{\label{tab:uvw} Galactic space velocities for the new sample of Li-rich giants reported in this work.}
\centering\small
\begin{tabular}{crrr}
\hline
\hline
CNAME & \multicolumn{1}{c}{U} & \multicolumn{1}{c}{V} & \multicolumn{1}{c}{W} \\
      &  \multicolumn{1}{c}{km s$^{-1}$} &  \multicolumn{1}{c}{km s$^{-1}$} &  \multicolumn{1}{c}{km s$^{-1}$} \\
\hline
 08405643-5308309  & $-$36.2$\pm$2.0  &     $-$48.4$\pm$0.3    & $-$10.5 $\pm$0.6   \\
  17522490-2927512 &  --      &      --      &         --    \\ 
  17531013-2932063  & $-$16.1$\pm$0.2   &  20.7$\pm$2.2   &  1.8$\pm$1.2   \\
  18181062-3246291 &  43.5$\pm$0.5   &  31.1$\pm$1.8   &  $-$38.3$\pm$2.8   \\
  18182698-3242584 &  36.1$\pm$0.4  &   $-$65.8$\pm$7.6  &   $-$5.6$\pm$1.2   \\
  18265248+0627259 &   102.5$\pm$7.6  &   $-$47.6$\pm$8.2   &  $-$66.3$\pm$9.0     \\
  19223053+0138518  & $-$11.5$\pm$0.4  &   $-$10.7$\pm$0.5   &  $-$19.0$\pm$2.2   \\
  19251759+0053140  & 111.4$\pm$3.2  &   20.2$\pm$3.9   &  7.2$\pm$1.4   \\
  19261134+0051569  & 125.2$\pm$11.1  &  $-$92.3$\pm$13.9  &  21.6$\pm$2.4  \\ 
  19263808+0054441  & $-$32.2$\pm$0.4  &   $-$41.4$\pm$0.8   &  $-$21.2$\pm$2.4   \\
  19264134+0137595  & 98.6$\pm$6.0  &     $-$31.5$\pm$7.1  &   25.1$\pm$2.8   \\
  19264917-0027469   & --      &      --      &      --         \\
  19265013+0149070  &     --      &      --      &     --     \\
  19265193+0044004   & $-$15.8$\pm$1.6   &  12.5$\pm$1.6   &  $-$15.6$\pm$2.5   \\
  19270600+0134446  &      --      &     --      &      --     \\
  19270815+0017461  &      --      &      --      &      --     \\
  19273856+0024149  &      --      &      --      &      --     \\
  19274706+0023447  & 117.2$\pm$9.1   &  $-$54.6$\pm$11.3   & 20.1$\pm$2.7   \\
  19280508+0100139  & 171.3$\pm$15.2  &  $-$75.6$\pm$18.8  &  18.3$\pm$3.3   \\
  19283226+0033072  & 102.3$\pm$5.0    &   $-$34.9$\pm$6.5   &  $-$5.2$\pm$0.9  \\
\hline
\end{tabular}
\end{table}

\subsection{Stellar population analysis}

The second data release (DR2) of \emph{Gaia} has provided astrometric information for more than 1.3$\times$10$^9$ objects with unprecedented quality \citep{2016A&A...595A...1G,2018arXiv180409365G,2018arXiv180409366L}. We have cross-matched our list of Li-rich giants with \emph{Gaia} DR2 and obtained parallaxes and proper motions for all the stars (Table \ref{tab:gaia} in the appendix).

The relative uncertainty of the parallaxes has median of about 10\% and is lower than 25\% for most stars. For this preliminary kinematic analysis, we assumed that the stellar distance is the inverse of the parallax. This assumption should provide accurate results for most of our stars, but not for all of them. We refer to \citet{2018arXiv180409376L} for a discussion of the uncertainties, correlations, and limitations of the parallaxes. We considered only giants whose relative uncertainty of the parallaxes is lower
than 15\%.

The calculation of the heliocentric Galactic space-velocity components ($U$, $V$, and $W$) and respective uncertainties is based on the equations presented in \citet{1987AJ.....93..864J}. The components are in the right-hand system, meaning that $U$ is positive towards the Galactic centre, $V$ is positive towards the Galactic rotation, and $W$ is positive towards the Galactic north pole. For this calculation, we assumed that the local standard of rest rotates with $V_{LSR}$ = 220 km s$^{-1}$ and adopted the 3D solar motion of $(U, V, W)_\odot $= (+10.0, +5.2, +7.2) from \citet{1998MNRAS.298..387D}.

The results are displayed in the left panel of Fig. \ref{fig:grv} and are listed in Table \ref{tab:uvw}. The velocities of the Li-rich giants do not deviate significantly from the behaviour expected of stars in the Galactic disc. In the right panel, we include the plot of the [$\alpha$/Fe] ratio as a function of [Fe/H]. The value of [$\alpha/Fe$] is an average of abundances of [\ion{Mg}{i}/Fe], [\ion{Si}{i}/Fe], [\ion{Ca}{i}/Fe], [\ion{Ti}{i}/Fe], and [\ion{Ti}{ii}/Fe]. The Li-rich giants are compared to the field stars included in the iDR5 catalogue. In this last panel, the dots are results of Monte Carlo Markov chain (MCMC) simulations that take into account the uncertainties in the measured abundances. The contours in mark regions containing 95\% and 68\% of the data points. Using their [$\alpha$/Fe] ratios, we tentatively classify 11 giants as members of the thin disc and seven as members of the thick disc. No $\alpha$-element abundances are available for the remaining two giants. We note, however, that the tentative thin- and thick-disc giants do not separate well in the kinematic plot.

\begin{table*}
 \caption[]{\label{tab:logg} Seismic estimates of the surface gravity and stellar masses of the Li-rich giants in the CoRoT fields. }
\centering\small
\begin{tabular}{ccccccc}
\hline
\hline
CNAME & CoRoT ID & $\log~g$ & $\sigma_{\rm \log~g}$ & Mass & $\sigma_{\rm mass}$ & Num. \\
      &   &   &   & M$_{\odot}$ & M$_{\odot}$ & Pipelines \\
\hline
19223053+0138518 & 100440565 & 2.35 & 0.05 &  1.25 & 0.14 & 4\\
19251759+0053140 & 100919702 &  --  &  --  &  -- & -- & 0 \\
19261134+0051569 & 101064590 & 2.34 & 0.07 &  2.75 & 1.37 & 1 \\
19263808+0054441 & 101130864 & 2.48 & 0.02 &  0.91 & 0.12 & 2 \\   
19264134+0137595 & 101139596 & 2.39 & 0.07 &  1.55 & 0.19 & 3 \\   
19264917-0027469 & 101160340 & 1.93 & 0.06 &  1.46 & 0.46 & 1 \\   
19265013+0149070 & 101162874 &  --  &  --  &  -- & -- & 0 \\
19265193+0044004 & 101167637 & 2.42 & 0.03 &  1.20 & 0.15 & 4 \\   
19270600+0134446 & 101205220 & 2.48 & 0.05 &  1.74 & 0.56 & 1 \\   
19270815+0017461 & 101210895 & 2.23 & 0.02 &  1.29 & 0.18 & 1 \\   
19273856+0024149 & 101292381 & 2.10 & 0.04 &  1.54 & 0.27 & 3 \\   
19274706+0023447 & 101314825 &  --  &  --  &  -- & -- & 0 \\
19280508+0100139 & 101351658 & 2.30 & 0.08 &  3.03 & 1.60 & 1 \\   
19283226+0033072 & 101411079 &  --  &  --  &  -- & -- & 0 \\
\hline
\end{tabular}
\tablefoot{The error in $\log~g$ takes into account the error in $T_{\rm eff}$. Moreover, it includes both the standard and systematic errors for the targets for which detection was made by more than one pipeline. For the targets with only one determination, this is only the internal error of the pipeline (and thus it underestimates the uncertainty). The error in the masses also takes into account the error in $\nu_{\rm max}$, which is usually large. Moreover, we remark again that masses based on scaling relations need corrections that depend on the stellar parameters, as discussed in the text, and which were not applied here.}
\end{table*}

One of the tentative thick-disc stars, 18182698-3242584, was observed in a field towards the bulge, and thus it might instead be a member of this stellar population (but see Appendix \ref{sec:bulge} below), in particular, given the known chemical similarity between bulge and thick disc \citep{2010A&A...513A..35A}. Alternatively, it might also be a member of the $\alpha$-enhanced super-solar metallicity population identified by \citet{2011A&A...535L..11A}. A detailed discussion of the kinematic properties of the sample is beyond the scope of this paper, but we can conclude that they mostly seem to be disc giants.

\section{Discussion}\label{sec:disc}

In this section, we attempt to address the evolutionary stage of the stars in more detail. To do this, we make use of the CoRoT data available for a subsample of our Li-rich giants. We also make use of the recent \emph{Gaia} DR2 to compute luminosities and position the giants in the HR diagram. For the discussion in this section, we combine our sample of new discoveries with the \emph{Gaia}-ESO Li-rich giants previously reported in \citet{2016MNRAS.461.3336C}. For completeness, in the appendix we give both the observational data of these stars (Table~\ref{tab:obsold}) and the results of the re-analysis of their spectra in \emph{Gaia}-ESO iDR5 (Table~\ref{tab:parold}). Additional discussion of the Li-rich giants observed in the fields of open clusters is given in Appendix \ref{sec:clusters}. It is shown that these stars are not cluster members. An additional discussion of the giants observed towards the Bulge is given in Appendix \ref{sec:bulge}. It is also shown that Bulge membership is unlikely.

\subsection{Stellar properties from CoRoT data}\label{sec:seismic}

Perhaps the most important result based on the analysis of the CoRoT data\footnote{The CoRoT data are publicly available and can be downloaded at \url{http://idoc-corot.ias.u-psud.fr/}} of these new Li-rich giants is the evolutionary stage of star 19265193+0044004 (CoRoT 101167637). This star is a He-core burning clump giant according to  \citet{2011A&A...532A..86M}. This is one of the few known Li-rich giant with a clear asteroseismic determination of the evolutionary stage, and the first based on CoRoT data. The other such giants include the clump giant reported by \citet{2014ApJ...784L..16S}, the RGB bump giant reported by \citet{2015A&A...584L...3J}, and the two clump giants reported by \citet{2018arXiv180410955B}, all with Kepler data. We also highlight the Li-rich giant found by \citet{2014A&A...564L...6M} in one open cluster as its position at the CMD is consistent with the red clump. For the remaining Li-rich giants in our sample, the CoRoT data do not provide a clear evolutionary classification.
In Section \ref{sec:unidam}, we estimated the giant to have 1.3-1.4 $\pm$ 0.3 M$_{\odot}$. Thus, this is most likely a low-mass star, although the error bar does not exclude the possibility of an intermediate-mass value. As a low-mass star, it went through the He-core flash at the end of the RGB evolution. As suggested before \citep[e.g.][]{2011ApJ...730L..12K,2014A&A...564L...6M}, this episode must likely be related to the origin of the Li enrichment. Star 19265193+0044004 becomes now an important addition that
supports the connection between Li-rich giants with the clump evolution.

\begin{figure*}
\centering
\includegraphics[height = 7cm]{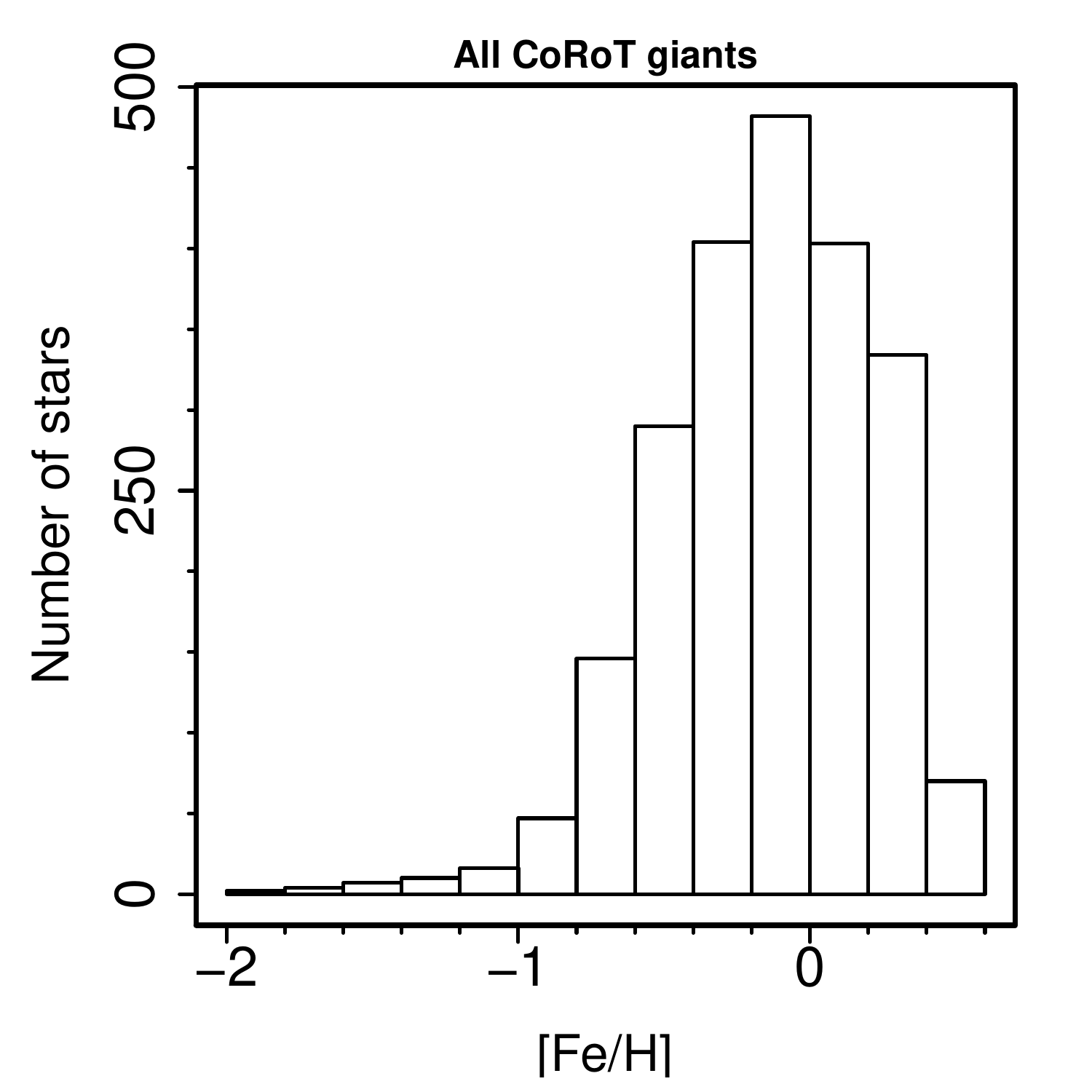}
\includegraphics[height = 7cm]{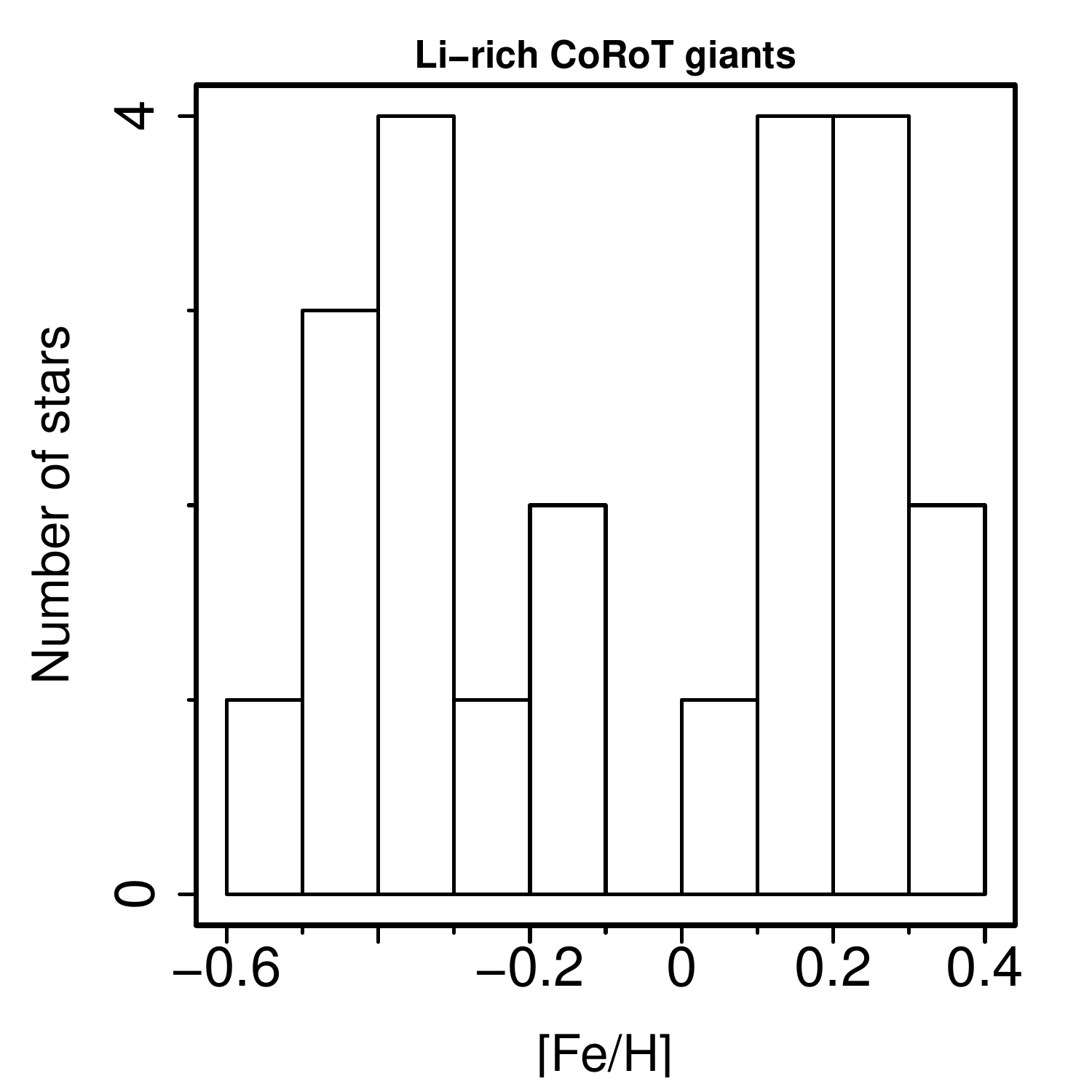}
\caption{Metallicity distribution of the CoRoT giants included in \emph{Gaia}-ESO iDR5. \emph{Left:}  All giants in the CoRoT fields with determinations of Li abundances. \emph{Right:} Li-rich giants.}\label{fig:hist}
\end{figure*}

Solar-like oscillations have been detected in the analysis of the light curves of 10 out of the 14 new Li-rich giants found in the CoRoT fields. Up to four pipelines were used in the seismic analysis \citep[see][]{2009A&A...508..877M,2010MNRAS.402.2049H,2010A&A...511A..46M,2018arXiv180504296D}. The frequency of maximum power, $\nu_{\rm max}$, together with our $T_{\rm eff}$ values was used to compute seismic estimates of $\log~g$ by means of a scaling relation \citep{1991ApJ...368..599B,1995A&A...293...87K}. Furthermore, using our temperatures, $\nu_{\rm max}$ , and the large separation ($\Delta\nu$), estimates of the stellar masses can also be obtained using scaling relations. See, for
instance, Eq. 3 of \citet{2012MNRAS.419.2077M}. We used the following solar values: $T_{\rm eff \odot}$ = 5777 K, $\nu_{\rm max \odot}$ = 3090 $\mu$Hz, and $\Delta\nu_{\odot}$ = 135 $\mu$Hz. The values are given in Table \ref{tab:logg}.

Masses derived using global seismic parameters and scaling relations  can be more accurate than those based on isochrones. However, the precision of the values obtained from the scaling relations itself depends on stellar parameters such as mass, metallicity, and evolutionary stage \citep[e.g.][]{2013EPJWC..4303004M}. Corrections based on theoretical models and frequencies are required to improve the precision of stellar mass values, such as in \citet{2017MNRAS.467.1433R} and Valentini et al. (in prep.). These corrections were not applied here. Therefore, the mass values should be seen only as indicative and used only as a check of the values derived previously using a different method. In most cases, the masses agree within the uncertainties with the values derived using {\sf UniDAM}. In some cases of large disagreement, the seismic mass is based on one detection, hence this can be seen as a difficult and uncertain measurement.

The seismic $\log~g$ values are mostly lower than the spectroscopic values. The mean difference is about $-$0.16 $\pm$ 0.13 dex. For the Li-rich CoRoT giants of higher metallicity, this change moves some of the stars from inside the region of the RGB bump to a position around the clump or closer to the early AGB. The
position of Li-rich CoRoT giants of lower metallicity is still consistent with the bump, although at higher stellar mass. Thus, the previous conclusion does not change. The stars remain consistent with three evolutionary stages: the RGB bump, the clump, and the early AGB.

Finally, \citet{2013A&A...555A..63D} analysed the CoRoT light curve of 19273856+0024149 and found it to display semi-sinusoidal variation, likely produced by rotation (we determined $v~\sin~i$ = 12 km s$^{-1}$). \citet{2013A&A...555A..63D} derived a variability period of 74.383 $\pm$ 1.0792 days. These two measurements yield
a ``projected" radius of the star of 17.6 R$_{\odot}$. Star 19273856+0024149 is the giant in the right panel of Fig.\ \ref{fig:newgiants} above the RGB bump above the blue track for 1.2 M$_{\odot}$. The models for 1.2 M$_{\odot}$ predict a radius of $\sim$ 11.6 R$_{\odot}$ for its $\log~g$ (2.39 dex, Table \ref{tab:par}). A larger radius might mean that the giant is brighter and/or has higher mass than implied by our spectroscopic $\log~g$ itself, bringing the star closer to the RGB bump. A higher mass value (1.54 M$_{\odot}$) is supported by the seismic analysis.

\begin{figure*}
\centering
\includegraphics[height = 14cm]{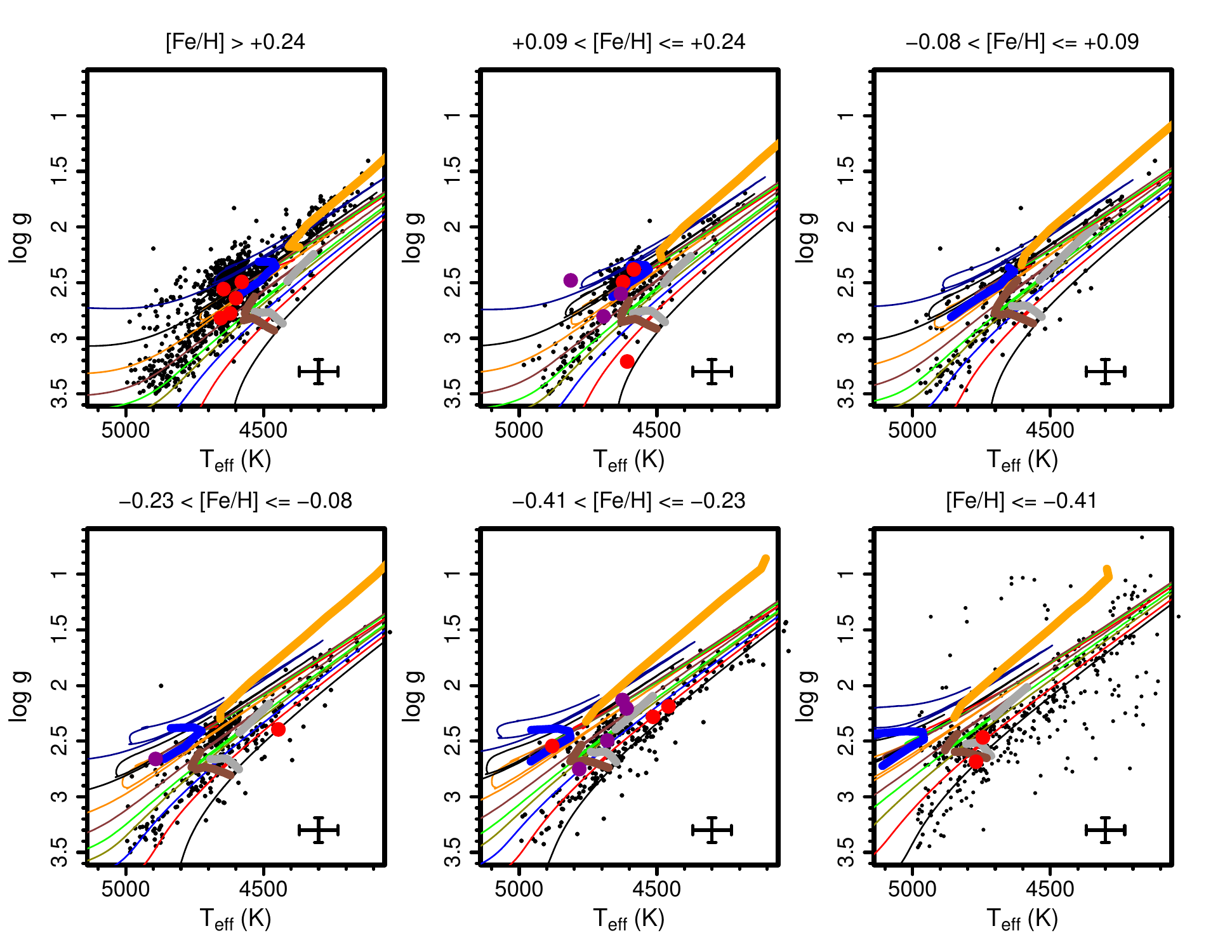}
\caption{All CoRoT giants with detected Li abundances or upper limits in the iDR5 catalogue. The giants found to be Li-rich are highlighted. The tracks are as those shown in Fig. \ref{fig:newgiants}. The red symbols are the new discoveries reported in this paper, while the dark magenta symbols are the Li-rich giants reported by \citet{2016MNRAS.461.3336C}.}\label{fig:allcorot}
\end{figure*}

\subsection{Giants in CoRoT fields}\label{sec:corot}

The extended sample of CoRoT giants indicates that evolutionary stage is one main factor that unites Li-rich giants. \emph{Gaia}-ESO has observed 2865 targets in CoRoT fields; 2431 of them have $\log~g$ $\leq$ 3.5 in the iDR5 catalogue (and thus are considered here to be giants). Lithium abundances or upper limits were derived for 2252 of them. 

The characteristics of the CoRoT data mean that the giants for which oscillations have been extracted are mostly in the range of the clump and bump, or somewhat lower on the RGB. These giants with asteroseismic data were the priority in the \emph{Gaia}-ESO observations. Nevertheless, because of the difficulties of assigning fibres during the observations, these giants were not observed
alone. The observed giants are distributed throughout the RGB, with $\log~g$ ranging from 3.5 to 0.70~dex (with quartiles 2.74 and 2.30~dex). Within this large sample, we found a total of 23 Li-rich giants, including those reported in this work and in \citet{2016MNRAS.461.3336C}. This is a fraction of 1.02\%, which is consistent with the numbers reported in the literature \citep[$\sim$1-2\%, see e.g.][]{1989ApJS...71..293B,2000AJ....119.2895P,2016ApJ...819..135K}.

Histograms with the metallicity distribution of the CoRoT giants are shown in Fig. \ref{fig:hist}. The metallicity distribution of all the CoRoT giants spans from [Fe/H] = $-$1.87 to +0.52, with mean $-$0.15$\pm$0.36 and quartiles at $-$0.38 and +0.13. The Li-rich CoRoT giants have mean [Fe/H] = $-$0.06$\pm$0.31 and the quartiles of the distribution are at $-$0.34 and +0.21. Thus, their metallicity distribution seems slightly shifted towards higher metallicities. The metallicity distribution is bimodal. There are two peaks, one at [Fe/H] = $-$0.40 (9 stars below $-$0.20) and another at +0.20 (11 stars above 0.0). Only two stars are found between [Fe/H] = $-$0.20 and 0.0 (and one more lacks determination of [Fe/H]). Given the small fraction of Li-rich giants, their numbers seems consistent with the metallicity distribution of the larger population.

The important result is that within this large sample of CoRoT giants, the Li-rich giants are mostly found in a narrow range of surface gravity values (i.e. narrow range of luminosities). These giants are mostly in the proximity of the RGB luminosity bump, although in particular for higher metallicity, some are also consistent with the position of the clump and/or could be at the early AGB. The asteroseismic data  classify one star as a red clump giant, and at least one giant is visually consistent with the core He-burning stage at the intermediate-mass regime, which might be even more challenging to explain. There is no extra-mixing event known to take place at this stage for intermediate-mass giants. The four exceptions likely have very uncertain atmospheric parameters. 

The concentration around the three evolutionary regions is clearly visible in Fig.\ \ref{fig:allcorot}, even though the error bars prevent an accurate positioning of the objects. This observation differs from previous reports that Li-rich giants are located throughout the whole extension of the RGB. For example, \citet{2011A&A...531L..12A} reported 1 Li-rich low-mass M-type giant likely at the tip of the RGB. \citet{2011A&A...529A..90M} discussed 5 Li-rich giants located between the RGB bump and the tip of the RGB. \citet{2013MNRAS.430..611M} reported 23 Li-rich giants distributed from the bottom to the tip of the RGB. 

This conclusion is also different from what has been reported in \citet{2016MNRAS.461.3336C}, where most giants had been found to lie below the position of the RGB luminosity bump. This concentration seemed to suggest planet engulfment as the most likely scenario. We note here again that the new evolutionary tracks of \citet{2018MNRAS.476..496F} argue for a lower position of the RGB bump of low-mass stars. For the sample of \citet{2016MNRAS.461.3336C}, we used a new set of parameters revised during the \emph{Gaia}-ESO iDR5 analysis. Most of the changes between the values of $T_{\rm eff}$ and $\log~g$ reported in \citet{2016MNRAS.461.3336C} and those reported here are well explained by the uncertainties in the measurements,
however. Moreover, the seismic $\log~g$ values place the stars closer to the red clump, making a position below the RGB bump even less likely.

Further motivation to review the likelihood of planet engulfment as a main channel behind the Li enrichment comes from recent works investigating Be abundances in Li-rich giants. \citet{2017PASJ...69...74T} attempted to detect Be in 20 Li-rich giants, including a few with Li abundances above the meteoritic value. No Be enhancement was detected. \citet{2018arXiv180104379A} recently reported an attempt to detect Be in two Li-rich giants. Again, the Be abundance was found to be depleted, as expected for red giants after the first dredge-up. Moreover, as discussed before, very fast rotation that clearly needs an additional mechanism to accelerate the stellar surface is seen in only two giants of our sample. The combination of all these observations seems to suggest that planet engulfment plays a minor role at most in the formation process of Li-rich giants.

\begin{figure*}
\centering
\includegraphics[height = 14cm]{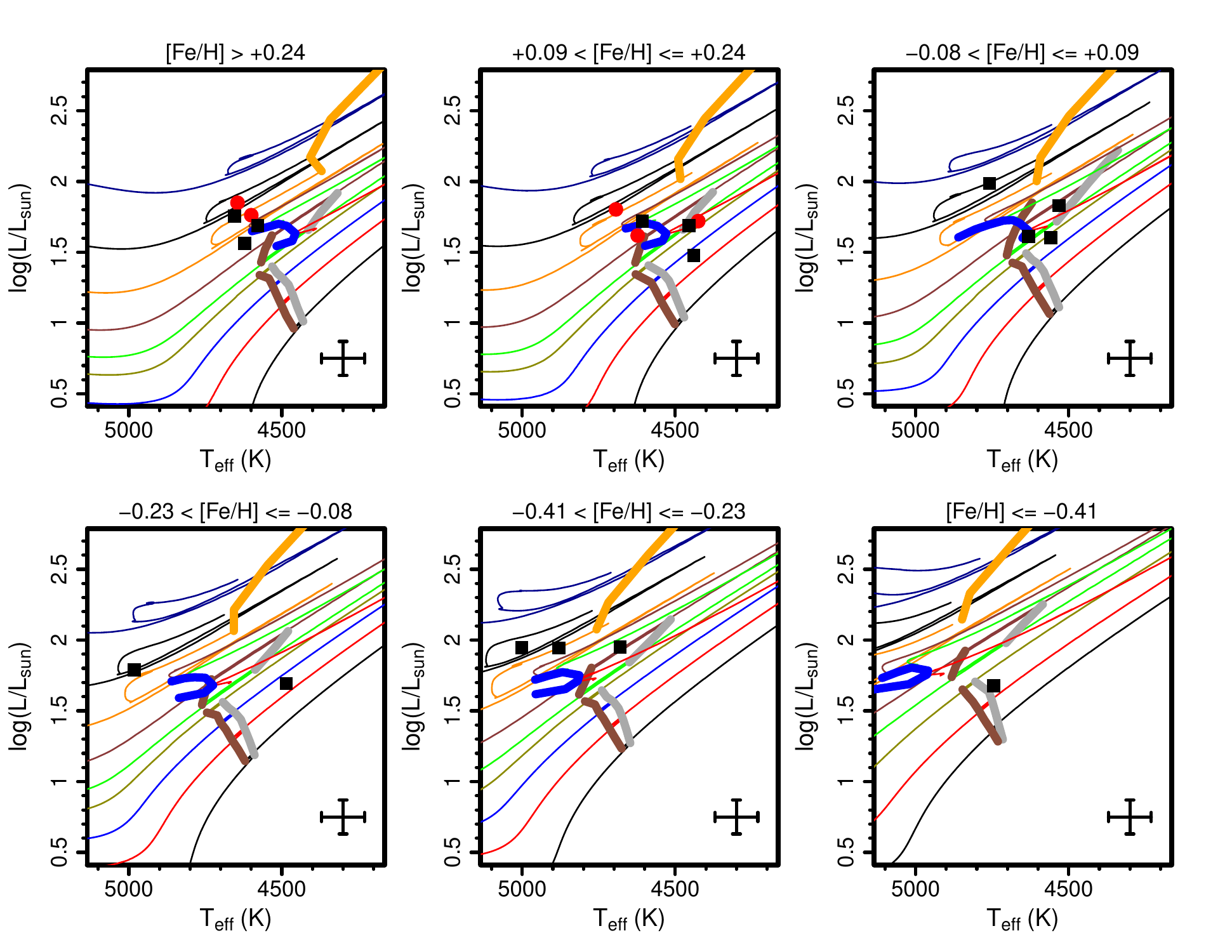}
\caption{HR diagram with 27 Li-rich giants discovered by the \emph{Gaia}-ESO Survey that have good values of \emph{Gaia} parallaxes, divided according to metallicity into different panels. The tracks are the same as in Fig. \ref{fig:newgiants}. The range of [Fe/H] is given at the top of each panel. Super Li-rich giants with A(Li) $>$ 3.3 dex (in non-LTE) are shown as red circles, and
giants with a Li abundance below this are shown as black squares. Typical error bars, $\pm$ 70 K in $T_{\rm eff}$ and $\pm$ 0.12 in $\log~(L/L_{\odot})$, are shown in the bottom right corners
of the panels.}\label{fig:lum}
\end{figure*}

\subsection{Luminosities with \emph{Gaia} DR2 data}\label{sec:gaia}

As mentioned before, the \emph{Gaia} DR2 parallaxes and proper motions are given in Table \ref{tab:gaia}. To calculate stellar luminosities, we assumed that the stellar distance is the inverse of the parallax. This assumption should provide accurate results for most of our stars, but not for all of them. We refer to \citet{2018arXiv180409376L} for a discussion of the uncertainties, correlations, and limitations of the parallaxes. We computed luminosities for the 27 giants whose relative uncertainty of the parallaxes is lower than 15\%.

To compute the absolute magnitudes, $Ks$ from 2MASS was transformed into $K$ in the CIT/CTIO system \citep{1982AJ.....87.1029E} using the relation from \citet{2001AJ....121.2851C} and ignoring the reddening. Bolometric corrections in $K$ have been tabulated by \citet{2000AJ....119.1448H}. The bolometric correction mostly
depends on the $T_{\rm eff}$ of the giant, is only weakly dependent on the metallicity, and is mostly independent of $\log~g$. Thus, in the grid of \citet{2000AJ....119.1448H}, we first selected the table of closer metallicity (either [Fe/H] = 0.00 or $-$0.50 dex) and linearly interpolated the values in $T_{\rm eff}$. Failing to interpolate in [Fe/H] causes an effect of at most 0.02 mag in the bolometric correction. This effect is negligible given that the uncertainties in our $T_{\rm eff}$ values can cause an effect of about 0.2 mag in the bolometric correction. Luminosities were computed using a solar bolometric magnitude of M$_{\rm bol~\odot}$ = 4.75 mag.

We estimated the uncertainties in the luminosities coming from the uncertainties in the parallaxes, $Ks$ magnitudes, and bolometric corrections. To do this, we assumed that the observed value is the mean of a Gaussian distribution with a standard deviation equal to the observed error (assumed to be 0.2 mag in the case of the bolometric correction). We then repeated the calculation of the luminosities 10\,000 times by drawing a random value of parallax, magnitude, and bolometric correction out of these distributions. The standard deviation of the resulting distribution of luminosity values was taken to be the uncertainty in this quantity. The median of the uncertainties in the luminosities is 0.12 dex.

The position in the HR diagram of the Li-rich giants for which we computed luminosities is shown in Fig. \ref{fig:lum}. The plot is divided into metallicity bins as in Fig. \ref{fig:newgiants} and shows the same evolutionary tracks. It is clear that the HR diagram shows a similar behaviour as the $T_{\rm eff}$-$\log~g$ diagram. The low-mass giants are concentrated around the position of the luminosity bump or the clump. The intermediate-mass stars seem to be at the core-He burning stage. A few stars might be at the early AGB, given the uncertainties.

Luminosity values of higher quality may be obtained with a better estimation of the distances and by taking into account the reddening. We note, however, that a reddening of 1 mag would increase the luminosity by 0.4 dex. This change in the luminosities would not change our conclusions. The giants would still be located around the bump, the core-He burning regions, or the early AGB. Thus, the Gaia data seem to confirm our previous conclusions.

As a final comment, we note that star 18265248+0627259, which seemed to be a core-He burning intermediate-mass star in Fig. \ref{fig:newgiants}, is in the same region in Fig \ref{fig:lum} (bottom left panel). This position makes internal mixing of fresh Li unlikely in this star. This fast rotator is also one of the candidates for planet engulfment. A follow-up investigation of its Be abundance would be an interesting way to search for additional support for this hypothesis.

\section{Summary}\label{sec:end}

We reported on the discovery of 20 new Li-rich giants observed by the \emph{Gaia}-ESO Survey. Four giants were observed in the field of open clusters, but do not seem to be members. Two giants were observed in fields towards the Galactic bulge, but magnitudes and proper motions are not compatible with bulge membership. The remaining 14 giants were observed in the CoRoT fields.

The asteroseismic data classify star 19265193+0044004 as a He-core burning clump giant. To the best of our knowledge, this is only the fifth Li-rich giant with asteroseismic determination of the evolutionary stage; it is the first with CoRoT data. It becomes the fourth such star to be found at the red clump. Its evolutionary stage supports a possible connection between the He-flash and the surface Li enrichment.

A comprehensive investigation of additional properties (IR magnitudes, rotational velocities, strong lines, and additional chemical abundances) did not reveal common peculiarities shared by all the Li-rich giants. We were able to identify one star with enhanced Ba abundance (19265013+0149070), five giants with $v~sin~i~>$ 10 km s$^{-1}$, and one star with IR excess (19270600+0134446). All giants show disc-like motion. Eleven stars seem to belong to the thin disc, and seven have enhanced [$\alpha$/Fe], which tentatively classifies them as thick-disc stars.

The two fastest rotators in our sample are candidates for having suffered planet engulfment. Otherwise, the only common characteristic of the Li-rich giants in our sample seems to be their evolutionary stage. The Li-rich giants are mostly located around three evolutionary stages: the RGB luminosity bump, the clump, and the early AGB. 

The concentration around these three evolutionary stages is particularly clear in the sample of giants in the CoRoT fields observed within the \emph{Gaia}-ESO Survey. Lithium abundances (or upper limits) are available for 2252 such giants, covering from the bottom to the upper regions of the RGB (for low-mass stars) and up to the early AGB of intermediate-mass stars. In this extended sample,
the 1\% of the giants that were found to be Li-rich are only located around the RGB luminosity bump, the clump, or the early
AGB. Luminosities computed using \emph{Gaia} DR2 parallaxes also support these conclusions.

This observation suggests that evolutionary stage plays a major role in the process of Li enrichment, at least in this sample. Additional processes such as planet engulfment probably only play a minor role. However, Li-rich objects found in other evolutionary stages cannot be explained in the same way. This includes the Li-rich metal-poor dwarfs and subgiants that were  found, for example, in the globular cluster NGC 6397 by \citet{2011ApJ...738L..29K}, in M 30 by \citet{2016A&A...589A..61G}, and in the field by \citet{2018ApJ...852L..31L}. Deep mixing that would add freshly synthesised Li is not possible in these stars. Thus, they likely require external pollution. Even in these cases, however, the non-detection of Be enhancement in the Li-rich dwarf in NGC 6397 argues against planet engulfment \citep{2014A&A...563A...3P}.

\begin{acknowledgements}
We thank the referee for the quick reports and for the useful suggestions that helped to improve the clarity of the paper. R. Smiljanic acknowledges support from the Polish Ministry of Science and Higher Education. G. Tautvai{\v s}ien{\. e} acknowledges support from the European Social Fund  via the Lithuanian Science Council grant No. 09.3.3-LMT-K-712-01-0103. V.A. and S.G.S acknowledge the support from Funda\c{c}\~ao para a Ci\^encia e a Tecnologia (FCT, Portugal) through the research grant through national funds and by FEDER through COMPETE2020 by grants UID/FIS/04434/2013 \& POCI-01-0145-FEDER-007672, PTDC/FIS-AST/1526/2014 \& POCI-01-0145-FEDER-016886 and PTDC/FIS-AST/7073/2014 \& POCI-01-0145-FEDER-016880. V.A. and S.G.S also acknowledge support from FCT through Investigador FCT contracts nr. IF/00650/2015/CP1273/CT0001 and IF/00028/2014/CP1215/CT0002. T.B. was funded by the project grant `The New Milky Way' from the Knut and Alice Wallenberg Foundation. JM  acknowledges  support from the ERC Consolidator Grant funding scheme (project STARKEY, G.A. n. 615604). T. Morel acknowledges financial support from Belspo for contract PRODEX GAIA-DPAC and PLATO. T. Masseron acknowledges support provided by the Spanish Ministry of Economy and Competitiveness (MINECO) under grant AYA-2017-88254-P. R.A.G acknowledges the support of the PLATO/CNES grant. S.M. acknowledges support from NASA grant NNX15AF13G and NSF grant AST-1411685 and the Ramon y Cajal fellowship no. RYC-2015-17697. AB acknowledges support from the Millennium Science Initiative (Chilean Ministry of Economy). The research leading to the presented results has received funding from the European Research Council under the European Community's Seventh Framework Programme (FP7/2007-2013) / ERC grant agreement no 338251 (StellarAges). Based on data products from observations made with ESO Telescopes at the La Silla Paranal Observatory under programme ID 188.B-3002. These data products have been processed by the Cambridge Astronomy Survey Unit (CASU) at the Institute of Astronomy, University of Cambridge, and by the FLAMES/UVES reduction team at INAF/Osservatorio Astrofisico di Arcetri. These data have been obtained from the \emph{Gaia}-ESO Survey Data Archive, prepared and hosted by the Wide Field Astronomy Unit, Institute for Astronomy, University of Edinburgh, which is funded by the UK Science and Technology Facilities Council. This work was partly supported by the European Union FP7 programme through ERC grant number 320360 and by the Leverhulme Trust through grant RPG-2012-541. We acknowledge the support from INAF and Ministero dell' Istruzione, dell' Universit\`a' e della Ricerca (MIUR) in the form of the grant "Premiale VLT 2012". The results presented here benefit from discussions held during the \emph{Gaia}-ESO workshops and conferences supported by the ESF (European Science Foundation) through the GREAT Research Network Programme. This research has made use of: NASA's Astrophysics Data System; the SIMBAD database, operated at CDS, Strasbourg, France; the VizieR catalogue access tool, CDS, Strasbourg, France. The original description of the VizieR service was published in \citet{2000A&AS..143...23O}; data products from the Wide-field Infrared Survey Explorer, which is a joint project of the University of California, Los Angeles, and the Jet Propulsion Laboratory/California Institute of Technology, funded by the National Aeronautics and Space Administration; data products from the Two Micron All Sky Survey, which is a joint project of the University of Massachusetts and the Infrared Processing and Analysis Center/California Institute of Technology, funded by the National Aeronautics and Space Administration and the National Science Foundation; the WEBDA database, operated at the Department of Theoretical Physics and Astrophysics of the Masaryk University. The analysis has made extensive use of {\sf R} \citep{Rcore}, {\sf RStudio} \citep{RStudio}, and the {\sf R} packages {\sf FITSio} \citep{FITSio}, {\sf gplots} \citep{gplots}, and {\sf stringr} \citep{stringr}. This work has made use of data from the European Space Agency (ESA) mission {\it Gaia} (\url{https://www.cosmos.esa.int/gaia}), processed by the {\it Gaia}
Data Processing and Analysis Consortium (DPAC, \url{https://www.cosmos.esa.int/web/gaia/dpac/consortium}). Funding for the DPAC has been provided by national institutions, in particular the institutions participating in the {\it Gaia} Multilateral Agreement.
\end{acknowledgements}

\bibliographystyle{aa} % style aa.bst
\bibliography{smiljanic} % your references Yourfile.bib

\begin{appendix}

\onecolumn 

\section{\emph{Gaia} DR2 data for the Li-rich giants discovered in the \emph{Gaia}-ESO Survey}

\begin{table*}[ht]
 \caption[]{\label{tab:gaia} \emph{Gaia} DR2 photometric and astrometric information for the complete list of 40 Li-rich giants discovered by the \emph{Gaia}-ESO Survey. The luminosities computed as discussed in the text are given in the last column. The new discoveries are listed in the top part of the table, and the stars from \citet{2016MNRAS.461.3336C} appear in the bottom part.}
\centering\small
\begin{tabular}{ccccccc}
\hline
\hline
CNAME & \emph{Gaia} DR2 Designation & $G$ & $\pi$ & pmRA & pmDEC & log(L/L$_{\odot}$)\\
 & & mag & mas & mas yr$^{-1}$ & mas yr$^{-1}$ &  \\
\hline
08405643-5308309 &  5318493981589528192 & 13.87 & 0.34$\pm$0.01 &  $-$2.70$\pm$0.03 &  +2.21$\pm$0.03 & 1.69$\pm$0.08 \\
17522490-2927512 &  4056552974204511104 & 14.93 & 0.28$\pm$0.07 &  $-$1.77$\pm$0.10 & $-$3.75$\pm$0.09  & -- \\
17531013-2932063 &  4056548065148248576 & 13.71 & 0.36$\pm$0.05 &   +1.00$\pm$0.07 &  +0.78$\pm$0.05  & -- \\
18181062-3246291 &  4045590259161801088 & 11.69 & 0.73$\pm$0.05 &   +7.33$\pm$0.08 &  +0.69$\pm$0.07  & 1.60$\pm$0.11 \\
18182698-3242584 &  4045596512634175232 & 12.58 & 0.47$\pm$0.05 &  $-$2.41$\pm$0.08 & $-$6.64$\pm$0.07 & 1.72$\pm$0.13 \\
18265248+0627259 &  4477215166550061184 & 14.04 & 0.26$\pm$0.03 &   +1.47$\pm$0.05 & $-$6.58$\pm$0.05 & 1.79$\pm$0.13\\
19223053+0138518 &  4288628856312126848 & 12.55 & 0.60$\pm$0.04 &   +3.30$\pm$0.08 & $-$1.63$\pm$0.08 & 1.69$\pm$0.10 \\
19251759+0053140 &  4263760067953345920 & 14.02 & 0.40$\pm$0.04 &  $-$2.75$\pm$0.07 & $-$3.28$\pm$0.06 & 1.56$\pm$0.11\\
19261134+0051569 &  4287730555301937664 & 14.50 & 0.25$\pm$0.03 &  $-$4.53$\pm$0.05 & $-$6.57$\pm$0.05 & 1.68$\pm$0.14\\
19263808+0054441 &  4287729691992814848 & 12.61 & 0.59$\pm$0.04 &   +3.38$\pm$0.07 & $-$3.33$\pm$0.06 & 1.75$\pm$0.10 \\
19264134+0137595 &  4287845866587261056 & 13.81 & 0.29$\pm$0.03 &  $-$3.61$\pm$0.06 & $-$3.84$\pm$0.05 & 1.85$\pm$0.13 \\
19264917-0027469 &  4263365618150379904 & 15.49 & 0.22$\pm$0.05 &  $-$3.79$\pm$0.08 & $-$6.22$\pm$0.06 & -- \\
19265013+0149070 &  4288600681325536128 & 15.28 & 0.16$\pm$0.04 &  $-$3.57$\pm$0.08 & $-$4.72$\pm$0.06 & --  \\
19265193+0044004 &  4287713169274054400 & 12.58 & 0.42$\pm$0.04 &   +2.81$\pm$0.07 &  +0.68$\pm$0.06 & 1.94$\pm$0.12 \\
19270600+0134446 &  4287841605979616128 & 14.32 & 0.19$\pm$0.04 &  $-$1.30$\pm$0.11 &  +0.33$\pm$0.08 & -- \\
19270815+0017461 &  4263486804951740160 & 14.92 & 0.14$\pm$0.04 &  $-$4.21$\pm$0.11 & $-$6.03$\pm$0.08 & -- \\
19273856+0024149 &  4263495016929225984 & 14.80 & 0.14$\pm$0.04 &  $-$2.77$\pm$0.07 & $-$6.01$\pm$0.05 & -- \\
19274706+0023447 &  4263483609496036864 & 14.44 & 0.27$\pm$0.03 &  $-$4.01$\pm$0.06 & $-$5.26$\pm$0.05 & 1.72$\pm$0.14 \\
19280508+0100139 &  4287748250568596608 & 14.90 & 0.22$\pm$0.03 &  $-$4.48$\pm$0.06 & $-$6.39$\pm$0.05 & 1.62$\pm$0.15 \\
19283226+0033072 &  4287505739532706048 & 14.21 & 0.27$\pm$0.02 &  $-$2.04$\pm$0.04 & $-$4.60$\pm$0.04 & 1.76$\pm$0.12 \\
\hline
08095783-4701385 &  5519275280241807744 & 12.27 & 0.44$\pm$0.03 & $-$10.11$\pm$0.07 &  +5.06$\pm$0.06 & 1.95$\pm$0.10 \\
08102116-4740125 &  5519174438707586560 & 13.74 & 0.22$\pm$0.02 &  $-$4.27$\pm$0.03 &  +4.86$\pm$0.04 & 2.07$\pm$0.11 \\
08110403-4852137 &  5516065363421606400 & 14.41 & 0.16$\pm$0.02 &  $-$3.18$\pm$0.04 &  +4.57$\pm$0.04 & 1.98$\pm$0.15 \\
08395152-5315159 &  5318117570655113472 & 14.92 & 0.19$\pm$0.03 &  $-$6.16$\pm$0.05 &  +6.44$\pm$0.04 & 1.83$\pm$0.14 \\
10300194-6321203 &  5252183088177166208 & 13.63 & 0.35$\pm$0.02 &  $-$9.74$\pm$0.03 &  +3.79$\pm$0.02 & 1.67$\pm$0.09 \\
10323205-6324012 &  5251997786110150016 & 13.31 & 0.44$\pm$0.01 & $-$12.34$\pm$0.03 &  +3.82$\pm$0.02 & 1.48$\pm$0.08 \\
10495719-6341212 &  5241195634131057024 & 13.26 & 0.42$\pm$0.01 &  $-$4.22$\pm$0.02 &  +2.15$\pm$0.02 & 1.69$\pm$0.09 \\
10503631-6512237 &  5239541728112201216 & 13.93 & 0.24$\pm$0.02 & $-$10.53$\pm$0.03 &  +3.90$\pm$0.03 & 1.93$\pm$0.11 \\
11000515-7623259 &  5201529682666727680 & 12.73 & 0.53$\pm$0.02 & $-$11.65$\pm$0.04 &  +3.50$\pm$0.03 & 1.89$\pm$0.09 \\
18033785-3009201 &  4050184607210512512 & 13.64 & 0.35$\pm$0.03 &   +0.36$\pm$0.07 & $-$2.10$\pm$0.05 & 1.69$\pm$0.11 \\
19230935+0123293 &  4264555358460629632 & 15.34 & 0.12$\pm$0.04 &  $-$5.28$\pm$0.06 & $-$5.82$\pm$0.06 & -- \\
19242472+0044106 &  4263749931831068928 & 13.78 & 0.42$\pm$0.03 &  $-$4.20$\pm$0.06 & $-$7.83$\pm$0.05 & 1.61$\pm$0.11 \\
19252571+0031444 &  4263697189630154112 & 14.94 & 0.18$\pm$0.04 &  $-$1.91$\pm$0.09 & $-$3.88$\pm$0.07 & --  \\
19252758+0153065 &  4288618754548267904 & 13.36 & 0.33$\pm$0.03 &  $-$2.44$\pm$0.06 & $-$3.08$\pm$0.05 & 1.80$\pm$0.13 \\
19252837+0027037 &  4263682689819998208 & 15.83 & 0.09$\pm$0.05 &  $-$3.67$\pm$0.08 & $-$3.61$\pm$0.07 & -- \\
19253819+0031094 &  4263685301160162048 & 15.25 & 0.16$\pm$0.04 &  $-$2.89$\pm$0.09 & $-$7.22$\pm$0.07 & -- \\
19261007-0010200 &  4263433753514379648 & 13.86 & 0.22$\pm$0.02 &  $-$3.38$\pm$0.04 & $-$2.06$\pm$0.04 & 1.95$\pm$0.12 \\
19264038-0019575 &  4263381629788846720 & 14.94 & 0.23$\pm$0.04 &  $-$3.32$\pm$0.06 & $-$9.02$\pm$0.05 & -- \\
19301883-0004175 &  4215367778069753728 & 14.13 & 0.15$\pm$0.04 &  $-$3.82$\pm$0.06 & $-$5.68$\pm$0.05 & --\\
19304281+2016107 &  2017726986620705152 & 14.53 & 0.25$\pm$0.04 &  $-$0.89$\pm$0.04 & $-$5.39$\pm$0.05 & 1.99$\pm$0.15 \\
\hline
\end{tabular}
\end{table*}

\pagebreak

\section{Tables with data for the Li-rich giants reported in \citet{2016MNRAS.461.3336C}.}

\begin{table*}[ht]
 \caption[]{\label{tab:obsold} Observational data of the Li-rich giants reported in \citet{2016MNRAS.461.3336C}.}
\centering\small
\begin{tabular}{cccccccr}
\hline
\hline
CNAME & Field & 2MASS ID & R.A. & DEC. & $V$ & RV & S/N \\
 & &  & h:m:s (J2000) & d:m:s (J2000) & mag & km s$^{-1}$ & \\
\hline
08095783-4701385 & $\gamma^2$ Vel & 08095784-4701383 & 08:09:57.83 & $-$47:01:38.5 & 10.82 & +26.1 & 145 \\
08102116-4740125 & $\gamma^2$ Vel & 08102116-4740125 & 08:10:21.16 & $-$47:40:12.5 & 14.22$^{1}$ & +70.7 & 129 \\
08110403-4852137 & NGC2547 & 08110403-4852137 & 08:11:04.03 & $-$48:52:13.7 & 14.87 & +54.2 &  58 \\
08395152-5315159 & IC2391 & 08395152-5315159 & 08:39:51.52 & $-$53:15:15.9 & 15.41 & +27.0 & 102 \\
10300194-6321203 & IC2602 & 10300194-6321203 & 10:30:01.94 & $-$63:21:20.3 & 14.16 & $-$10.4 & 145 \\
10323205-6324012 & IC2602 & 10323205-6324012 & 10:32:32.05 & $-$63:24:01.2 & 13.72 & +13.3 &  88 \\
10495719-6341212 & IC2602 & 10495719-6341212 & 10:49:57.19 & $-$63:41:21.2 & 13.84$^{1}$ & +13.8 & 123 \\
10503631-6512237 & IC2602 & 10503632-6512237 & 10:50:36.31 & $-$65:12:23.7 & 12.77 & $-$34.1 &  84 \\
11000515-7623259 & Cha I & 11000515-7623259 & 11:00:05.15 & $-$76:23:25.9 & 13.74$^{1}$ & $-$15.8 & 103 \\
18033785-3009201 & Bulge & 18033785-3009200 & 18:03:37.85 & $-$30:09:20.1 & 13.27$^{1}$ & $-$69.9 &  97 \\
19230935+0123293 & Corot & 19230934+0123293 & 19:23:09.35 & +01:23:29.3 & 15.93 & +11.9 &   7 \\
19242472+0044106 & Corot & 19242474+0044104 & 19:24:24.73 & +00:44:10.5 & 14.45 & +77.7 &  92 \\
19252571+0031444 & Corot & 19252571+0031444 & 19:25:25.71 & +00:31:44.4 & 15.10$^{1}$  & $-$38.5 &  44 \\
19252758+0153065 & Corot & 19252758+0153064 & 19:25:27.58 & +01:53:06.5 & 13.73 & +28.1 &  35 \\
19252837+0027037 & Corot & 19252837+0027037 & 19:25:28.37 & +00:27:03.7 & 16.02$^{1}$  & +0.1 &  28 \\
19253819+0031094 & Corot & 19253819+0031094 & 19:25:38.19 & +00:31:09.4 & 15.59 & +26.4 &  33 \\
19261007-0010200 & Corot & 19261020+0010226 & 19:26:10.07 & +00:10:20.0 & 14.55$^{1}$ & $-$21.1 &  63 \\
19264038-0019575 & Corot & -- & 19:26:40.38 & +00:19:57.5 & -- & +42.3 &  21 \\
19301883-0004175 & Corot & -- & 19:30:18.83 & +00:04:17.5 & -- & +57.3 &  41 \\
19304281+2016107 & NGC6802 & 19304281+2016107 & 19:30:42.81 & +20:16:10.7 & 14.67$^{3}$ & +17.4 &  67 \\
\hline
\end{tabular}
\tablefoot{The $V$ magnitudes are from APASS \citep[the AAVSO Photometric All-Sky Survey,][]{2015AAS...22533616H} unless otherwise noted: (1) NOMAD catalogue \citep{2004AAS...205.4815Z}; (2) The Guide Star Catalog, Version 2.3.2 (GSC2.3) (STScI, 2006).}
\end{table*}

\begin{table*}[ht]
 \caption[]{\label{tab:parold} New iDR5 atmospheric parameters and lithium abundances for the Li-rich giants reported in \citet{2016MNRAS.461.3336C}.}
\centering\small
\begin{tabular}{cccccccccccc}
\hline
\hline
CNAME & T$_{\rm eff}$ & $\sigma$ & $\log~g$ & $\sigma$ & [Fe/H] & $\sigma$ & $\xi$ & $\sigma$ & A(Li) & $\sigma$ & A(Li)\\
            &   (K) & (K) &  &  &  &  & km s$^{-1}$ & km s$^{-1}$ & (LTE)  & & (non-LTE) \\
\hline 
08095783-4701385 & 5002 &  28 & 2.55 & 0.04 & $-$0.25 & 0.02 & 1.50 & 0.00 & 3.60 &   -- & 3.24 \\
08102116-4740125 & 4433 & 175 &   -- &   -- & $-$0.12 & 0.01 &   -- &   -- & 3.52 & 0.11 &   -- \\
08110403-4852137 & 4599 & 212 &   -- &   -- & $-$0.07 & 0.08 &   -- &   -- &   3.60$^{1}$ &   0.13 &   -- \\
08395152-5315159 & 4531 & 187 & 2.52 & 0.18 &  +0.01 & 0.05 &   -- &   -- &   1.88$^{1}$ &   0.29 &  2.07 \\
10300194-6321203 & 4472 & 184 &   -- &   -- & $-$0.03 & 0.04 &   -- &   -- &   2.89$^{1}$ &   0.28 &   -- \\
10323205-6324012 & 4440 & 178 & 2.57 & 0.19 &  +0.15 & 0.01 &   -- &   -- &   2.96$^{1}$ &   0.23 &  2.89 \\
10495719-6341212 & 4646 & 117 &   -- &   -- &  +0.02 & 0.09 &   -- &   -- &   2.97$^{1}$ &  0.20 &   -- \\
10503631-6512237 & 4580 & 119 &   -- &   -- & $-$0.03 & 0.05 &   -- &   -- & 2.50 & 0.17 &   -- \\
11000515-7623259 & 4418 &  -- &   -- &   -- &  +0.14 & 0.00 &   -- &   -- & 2.55 & 0.07 &   -- \\
18033785-3009201 & 4455 &  58 & 2.43 & 0.11 &  +0.12 & 0.07 & 1.64 & 0.18 & 2.66 & 0.13 & 2.67 \\
19230935+0123293 & 4610 & 144 & 2.21 & 0.21 & $-$0.33 & 0.55 &   -- &   -- & 2.59 & 0.15 & 2.57 \\
19242472+0044106 & 4631 &  58 & 2.60 & 0.12 &  +0.09 & 0.06 & 1.73 & 0.20 & 2.51 & 0.06 & 2.57 \\
19252571+0031444 & 4892 & 169 & 2.66 & 0.25 & $-$0.17 & 0.20 &   -- &   -- & 2.32 & 0.11 & 2.38 \\
19252758+0153065 & 4694 &  46 & 2.80 & 0.10 &  +0.22 & 0.26 &   -- &   -- & 3.54 & 0.07 & 3.35 \\
19252837+0027037 & 4813 & 236 & 2.48 & 0.19 &  +0.14 & 0.24 &   -- &   -- & 3.32 & 0.13 & 3.17 \\
19253819+0031094 & 4625 & 247 & 2.13 & 0.41 & $-$0.40 & 0.27 &   -- &   -- & 3.28 & 0.07 & 3.06 \\
19261007-0010200 & 4680 &  35 & 2.49 & 0.09 & $-$0.35 & 0.18 &   -- &   -- & 3.26 & 0.06 & 3.02 \\
19264038-0019575 & 4782 &  46 & 2.75 & 0.10 & $-$0.40 & 0.28 &   -- &   -- & 3.69 & 0.16 & 3.34 \\
19301883-0004175 & 4128 &  77 & 1.22 & 0.31 &    -- &   -- &   -- &   -- & 2.14 & 0.17 & 2.13 \\
19304281+2016107 & 4759 &  67 & 2.63 & 0.12 & $-$0.04 & 0.11 & 1.80 & 0.09 & 2.62 & 0.06 & 2.63 \\
\hline
\end{tabular}
\tablefoot{(1) The values of Li abundance for these five stars are not part of the final iDR5 Gaia-ESO catalogue. The values are missing from the main catalogue, likely because abundance measurements performed by different pipelines disagreed by a large margin. We report here instead the Li abundances rederived by only one of these pipelines, that of the Arcetri node \citep[see description of this analysis node in][]{2015A&A...576A..80L}. }
\end{table*}

\pagebreak

%\twocolumn

\section{Giants in open cluster fields}\label{sec:clusters}

We have identified four Li-rich giants in the field of three open clusters. If the stars are indeed members of the clusters, we could use the known cluster distances and reddening values to accurately position the objects in a CMD. This would allow a more robust understanding of their evolutionary stage than is possible with the spectroscopic diagram of Fig. \ref{fig:newgiants} \citep[see, e.g. the clump giant found by][in the open cluster Trumpler 5]{2014A&A...564L...6M}. Often, however, radial velocities and/or metallicities indicate that the Li-rich giants seem to be just field stars that are unrelated to the cluster \citep[see ][for recent examples]{2017MNRAS.469.1330A,2017A&A...602A..33F}. 

The four giants reported here do not seem to be cluster members. This was also the case of the Li-rich giants in open cluster fields found by \citet{2016MNRAS.461.3336C}.

Membership is excluded based on discrepant RVs. Star CNAME 08405643-5308309 has an RV = +55 km s$^{-1}$ , while the open cluster IC 2391 has a mean RV $\sim$ 15 km s$^{-1}$ \citep[based on five stars reported in][]{2017A&A...601A..70S}. Stars CNAME 17522490-2927512 and 17531013-2932063 have +81.6 and $-$25.8 km s$^{-1}$, respectively, while the cluster Rup 134 has a mean RV $\sim$ $-$41 km s$^{-1}$ (Magrini et al., in prep.). Star CNAME 18265248+0627259 (the fastest rotator in the sample) has an RV = +32.7 km s$^{-1}$ , while NGC 6633 has a mean RV $\sim$ $-$29 km s$^{-1}$ \citep{2017A&A...603A...2M}.

Six giants from \citet{2016MNRAS.461.3336C}, all observed in the field of open clusters, are missing $\log~g$ values in the iDR5 catalogue (Table \ref{tab:parold}). This happened because the disagreement between the two pipelines deriving $\log~g$ values for these stars increased in the new analysis cycle \citep[the pipelines are described in][]{2015A&A...576A..80L}. These values were thus considered unreliable and discarded during the homogenisation stage. To have an indicative value of $\log~g$, we retrieved the values of one of these pipelines (the one that remained more consistent between the different analysis cycles). The $\log~g$ values for these giants are between 2.56 and 2.78. The $T_{\rm eff}$ values are between 4400 and 4650 K (Table \ref{tab:parold}). We verified that these values place the giants exactly around the RGB bump of low-mass stars in the right panel of Fig. \ref{fig:newgiants}. Thus, they would still support the conclusion drawn from the remaining stars.

\section{Giants towards the bulge}\label{sec:bulge}

\citet{2016MNRAS.461.3336C} reported the discovery of one Li-rich giant towards the bulge (CNAME 18033785-3009200), which seemed to have properties (RV and abundances) consistent with bulge membership. Here, we report on two additional Li-rich giants observed in fields towards the bulge (CNAME 18181062-3246291 and 18182698-3242584). 

Similarly as with open clusters, if we can confirm their membership to the bulge, we could use the known bulge distance and reddening maps \citep[e.g.][]{2013ApJ...769...88N} to accurately position the objects in a CMD.

One of our two new Li-rich giants, 18182698-3242584, has the highest Li enrichment discovered so far in the \emph{Gaia}-ESO Survey, with A(Li) = 4.04 in non-LTE. 

The two new giants have been included in a few proper motion studies dedicated to bulge fields \citep{2007AJ....134.1432V,2011A&A...534A..91T}. In particular, \citet{2007AJ....134.1432V} discussed the distribution of proper motions of bulge stars at Plaut's low extinction window; ($l$,$b$) = (0$^{\circ}$,8$^{\circ}$). In this field, the distributions of proper motions of bulge stars in Galactic coordinates peak at ($\mu_b$, $\mu_{l}\cos~b$) $\sim$ (0, $-$2) mas yr$^{-1}$ (see their Fig. 8). The proper-motion dispersion of the bulge stars is found to be (3.39$\pm$0.11, 2.91$\pm$0.09). The proper motions derived by \citet{2007AJ....134.1432V} for stars 18181062-3246291 and 18182698-3242584 are ($\mu_b$, $\mu_{l}\cos~b$) = (6.8$\pm$0.7,$-$5.9$\pm$0.7) and ($-$7.5$\pm$0.04, $-$1.5$\pm$0.3), respectively. The values are only marginally consistent with the typical bulge values, suggesting that the giants are probably not bulge stars.

The two new Li-rich giants are part of fields observed by the VVV survey \citep[Vista Variables in the V{\'{\i}}a L\'actea,][]{2012A&A...537A.107S,2017yCat.2348....0M}, but when measurements are given, all magnitudes are flagged as unreliable (probably because of saturation). The unreliable $Ks$ value for 18181062-3246291 is 9.21 mag, which would be much brighter than the typical $Ks$ magnitude of bulge clump giants in the same field, $Ks$ $\sim$ 12.7-13.4 \citep{2012A&A...544A.147S}.

Star 18033785-3009200, reported in \citet{2016MNRAS.461.3336C}, has been observed by OGLE \citep[Optical Gravitational Lensing Experiment,][]{2002AcA....52..217U} and has magnitudes $I$ = 12.63 mag and $V$ = 14.18 mag. Its VVV $Ks$ = 11.07 mag is this time more reliable. In both cases, the magnitudes again seem to be too bright for bulge giants \citep[see, e.g.][]{2004MNRAS.349..193S,2013ApJ...769...88N}.

Moreover, the distance we derived using {\sf UniDAM} for star 18181062-3246291 and the distances based on \emph{Gaia} parallaxes are too small to be consistent with bulge membership. We thus conclude that most likely none of the three \emph{Gaia}-ESO Li-rich giants observed in bulge fields belongs to the bulge itself.

\end{appendix}

\end{document}